\documentclass[eqsecnum,nofootinbib,11pt]{revtex4}
\usepackage{amsmath}
   \usepackage{bm}
\usepackage{amsthm,amscd}
\usepackage{mathrsfs}
\usepackage{verbatim}
\usepackage{amsfonts,amsmath,latexsym,amssymb,txfonts,bm}
\usepackage[american]{babel}
\usepackage{dsfont}
\usepackage[dvips]{graphicx}
\usepackage{subfigure}

\newcommand\bea{\begin{eqnarray}}
\newcommand\eea{\end{eqnarray}}

\begin{document}
\title{Conical Casimir Pistons with Hybrid Boundary Conditions}
\author{
Guglielmo Fucci\footnote{Electronic address: Guglielmo\textunderscore Fucci@Baylor.edu} and Klaus Kirsten\footnote{Electronic address: Klaus\textunderscore Kirsten@Baylor.edu}
\thanks{Electronic address: gfucci@nmt.edu}}
\affiliation{Department of Mathematics, Baylor University, One Bear Place 9653, Waco TX 76798 USA
}
\date{\today}
\vspace{2cm}
\begin{abstract}

In this paper we compute the Casimir energy and force for massless scalar fields endowed with hybrid boundary conditions, in the setting of the bounded generalized cone.
By using spectral zeta function regularization methods, we obtain explicit expressions for the Casimir energy and force in arbitrary dimensions in terms of the zeta function defined on the piston.
Our general formulas are, subsequently, specialized to the case in which the piston is modelled by a $d$-dimensional sphere. In this particular situation, explicit
results are given for $d=2,3,4,5$.

\end{abstract}
\maketitle

\section{Introduction}

The study of Casimir energy and force has become a major field of investigation in the past several years starting with the seminal
work \cite{casimir48} where this phenomenon was first predicted. Calculations of Casimir energy have been performed for a
plethora of different geometric configurations and boundary conditions (see e.g. \cite{bordag01,bordag09,milton01,plunien86}) leading to a variety
of interesting results. It is well known, however, that the evaluation of Casimir energies is plagued with divergences which need to be regularized
and renormalized \cite{blau88,bordag01,bordag09,bytsenko03,elizalde,elizalde94,milton01}. This is one of the reasons which lies at the roots of the increased interest in
Casimir piston configurations introduced a few years ago in \cite{cavalcanti04}. In fact, Casimr pistons often allow for an unambiguous
prediction of the force. One of the most interesting aspects of problems related to pistons is the determination of the sign of the resulting force.
Although one cannot predict a priori the sign of the force for an arbitrary configuration, several calculations have shown that it
depends critically not only on the boundary conditions imposed, but also on the specific geometry and topology of the system under consideration.
For rectangular Casimir pistons endowed with the same type of boundary conditions on all its sides, the Casimir force is such that
the piston is always attracted to the closest wall \cite{edery07,hertz05,hertz07}. The same kind of conclusions have been reached also for
more general Casimir pistons modelled by a compactly supported potential in presence of compactified extra-dimensions \cite{morales10,kirsten09}
and for two identical plates inside a cylinder \cite{marachevsky07}.

A repulsive Casimir force is obtained by considering two plates associated with different boundary conditions (Dirichlet on one and Neumann on the other)
\cite{boyer74}. Moreover, similar conclusions can be drawn for spherical shells in which the Casimir energy tends to generate a pressure that increases
its radius, invalidating in this way the Casimir model for the stability of the electron \cite{boyer68}.
Casimir piston systems leading to a repulsive force have been analyzed, for instance, in \cite{barton06}, \cite{fulling07} and \cite{li97}.
Because of this interesting behavior, Casimir energy and force for hybrid boundary conditions have attracted, in recent times, increased
interest (see e.g. \cite{zhai07} and references therein).

In this work, which represents a continuation
of the investigations started in \cite{fucci11}, we analyze conical Casimir pistons endowed with hybrid boundary conditions. Conical pistons
represent a generalization of the standard piston geometries in which both chambers have, essentially, the same geometry.
In fact, in the case of a conical piston, the most interesting feature is that one of the chambers contains a conical singularity at
the origin while the other does not. Because of this characteristic, it is of particular interest
to understand the behavior of the Casimir force when the piston approaches the singular point. Another type of Casimir piston constructed from two adjacent lunes
separated by a hemispherical piston has been recently considered in \cite{dowker11}.

The main physical motivation for the study of conical manifolds resides in models of quantum field theory requiring orbifold compactification \cite{fucci10,vongersdorff08}.
In fact an orbifold is defined, locally, by the quotient space of a smooth manifold $X$ and a discrete isometry group $G$. The action of the group on the
manifold possesses, in the general case, fixed points which are mapped to conical singularities in the quotient space. These models are a topic of great interest especially in the framework
of string theory \cite{bailin99,bezerra}.

In this paper we utilize $\zeta$-function methods in order to arrive at explicit expressions for the Casimir energy of
the conical piston. We denote by $\lambda_{n}$, with $n\in \mathbb{N}^{+}$, the spectrum of a self-adjoint partial differential operator acting on square integrable functions defined on smooth compact
manifolds. The spectral $\zeta$-function associated with the operator is then defined as
\begin{equation}\label{0}
\zeta(s)=\sum_{n=1}^\infty\lambda_{n}^{-s}\;,
\end{equation}
which is convergent for $\Re(s)>D/2$, with $D$ being the dimension of the manifold under consideration.
One can analytically continue, in a unique way, $\zeta(s)$ to a meromorphic function possessing only simple poles in the whole
complex plane which coincides with (\ref{0}) in its domain of convergence.

The outline of the paper is as follows. In the next section we delineate the geometry of the conical piston and we introduce the basic objects needed for our study.
In the framework of $\zeta$-function regularization, we obtain
specific expressions for the Casimir force, with hybrid boundary conditions, which manifest their dependence on the particular geometry of the piston. As a special case of our general
formulas we consider the situation in which the piston is a $d$-dimensional sphere and give very explicit results for low dimensional cases, namely $d=2,3,4,5$.
The conclusions stress the most important results of the article.

\section{The Geometry of the Conical Piston}

We will base our analysis on the bounded generalized cone which is a $D=(d+1)$-dimensional manifold defined as the direct product $\mathscr{M}=I\times\mathscr{N}$.
Here, $\mathscr{N}$ is the base manifold, assumed to be a smooth Riemannian manifold possibly with boundary, and $I=[0,1]\subset \mathds{R}$.
The manifold $\mathscr{M}$ is locally described by the line element \cite{cheeger83}
\begin{equation}\label{1}
ds^{2}=dr^{2}+r^{2}d\Sigma_{\mathscr{N}}^{2}\;,
\end{equation}
where $d\Sigma_{\mathscr{N}}^{2}$ represents the metric on $\mathscr{N}$ and $r\in I$.
For this type of singular Riemannian manifold the heat kernel and functional determinant of the
associated Laplace operator have been studied for massless and massive fields in \cite{bordag96,fucci10}.

The conical piston configuration that is associated to the generalized cone has been described in detail
in \cite{fucci11}. Let $a\in(0,b)$, and $\mathscr{N}_{a}$ be the associated cross section of the generalized cone $\mathscr{M}$
at the position $r=a$. This cross section divides the manifold $\mathscr{M}$ into two different regions; $M_{I}=[0,a]\times \mathscr{N}$ and
$M_{II}=(a,b]\times \mathscr{N}$. The two regions (or chambers), joined at their common boundary, $\mathscr{N}_{a}$, constitute the
conical piston, where the piston itself is modelled by the $d$-dimensional manifold $\mathscr{N}_{a}$ \cite{fucci11}. Obviously,
the two regions have essentially a different geometric structure since $M_{I}$ contains the conical singularity at the origin while $M_{II}$ does not.

In this work, we will consider the Laplace operator $\Delta_{\mathscr{M}}$ acting on the Hilbert space $\mathscr{L}^{2}(\mathscr{M})$
of scalar functions on the generalized cone $\mathscr{M}$. The eigenvalues $\alpha_{i}$ of the operator under consideration are found
by imposing the equation
\begin{equation}\label{2}
\left(-\Delta_{\mathscr{M}}+m^{2}\right)\varphi_{i}=\alpha_{i}^{2}\varphi_{i}\;,
\end{equation}
where the mass $m$ has been introduced, at this point, for technical reasons but the limit $m\to 0$ will be taken in the final results. In the coordinates
used to write the line element (\ref{1}) the above eigenvalue problem takes the form
\begin{equation}\label{3}
\left(-\frac{\partial^{2}}{\partial r^{2}}-\frac{d}{r}\frac{\partial}{\partial r}-\frac{1}{r^{2}}\Delta_{\mathscr{N}}+m^{2}\right)\varphi_{i}=\alpha_{i}^{2}\varphi_{i}\;,
\end{equation}
with $\Delta_\mathscr{N}$ representing the Laplace operator on the manifold $\mathscr{N}$. The idea is to solve (\ref{3})
in the two regions separately obtaining, in this way, two sets of eigenvalues \cite{fucci11}. In the region containing the conical singularity, namely region $I$, the solution
which is regular at $r=0$ is
\begin{equation}\label{4}
\varphi_{I}=r^{\frac{1-d}{2}}J_{\nu}(\gamma_{I}r )\Phi(\Omega)\;,
\end{equation}
while in region $II$ the eigenfunctions are a linear combination of Bessel functions of first and second kind as
follows
\begin{equation}\label{5}
\varphi_{II}=r^{\frac{1-d}{2}}\Big[A\,J_{\nu}(\gamma_{II}r)+B\,Y_{\nu}(\gamma_{II}r)\Big]\Phi(\Omega)\;.
\end{equation}
In the above equations we have defined $\alpha^{2}_{j}=\gamma_{j}^{2}+m^{2}$, with $j=(I, II)$, to be the eigenvalues, respectively, in region $I$ and
region $II$, and $\Phi(\Omega)$ are angular functions on $\mathscr{N}$ satisfying the eigenvalue problem
\begin{equation}\label{6}
\Delta_{\mathscr{N}}\Phi(\Omega)=-\lambda^{2}\Phi(\Omega)\;.
\end{equation}
In addition, the index $\nu$ identifying the Bessel functions in the solutions (\ref{4}) and (\ref{5}) can be found to be
\begin{equation}\label{7}
\nu^{2}=\lambda^{2}+\frac{(1-d)^{2}}{4}\;.
\end{equation}

The basic object needed in our study is the spectral $\zeta$-function associated with the eigenvalue problems
in both regions $I$ and $II$. It is defined as
\begin{equation}\label{9}
\zeta_{j}(s)=\sum_{\gamma_{i}}(\gamma_{i}^{2}+m^{2})^{-s}\;,
\end{equation}
where $j$ represents either $I$ or $II$. The spectral $\zeta$-function in (\ref{9}) is convergent for $\Re(s)>D/2$ and is defined by analytic continuation
in the rest of the complex plane where it will present at most simple poles \cite{bytsenko03,elizalde,elizalde94,kirsten01}.
The total $\zeta$-function associated with the generalized cone $\mathscr{M}$, that is the relevant one for the computation of the
Casimir energy and force, is obtained as a sum of the $\zeta$-functions of the two regions
\begin{equation}\label{10}
\zeta_{\mathscr{M}}(s)=\zeta_{I}(s)+\zeta_{II}(s)\;.
\end{equation}
In order to keep the manifold $\mathscr{N}$ that represents the piston arbitrary, the spectral $\zeta$-functions $\zeta_{I}(s)$
and $\zeta_{II}(s)$ will be expressed in terms of the auxiliary function $\zeta_{\mathscr{N}}$
defined as \cite{bordag96,cheeger83}
\begin{equation}\label{11}
\zeta_{\mathscr{N}}(s)=\sum_{\nu} d(\nu)\nu^{-2s}\;,
\end{equation}
where $d(\nu)$ denotes the degeneracy of the scalar harmonics $\Phi(\Omega)$ on $\mathscr{N}$.

Let us focus our attention on the explicit evaluation of the Casimir energy for the conical piston. It is well known
that the Casimir energy is defined in terms of the $\zeta$-function as follows \cite{bordag01,bordag09,elizalde,elizalde94,kirsten01}
\begin{equation}\label{0a}
E_{\textrm{Cas}}=\lim_{\alpha\to 0}\frac{\mu^{2\alpha}}{2}\zeta_{\mathscr{M}}\left(\alpha-\frac{1}{2}\right)\;,
\end{equation}
where $\mu$ represents an arbitrary parameter with the dimensions of a mass. Let us stress that in the
process of analytic continuation $\zeta_{\mathscr{M}}$ develops a simple pole at $s=-1/2$
being described by a Laurent expansion of the form \cite{bytsenko03,bordag01,bordag09,elizalde,elizalde94,kirsten01}
\begin{equation}\label{12}
 \zeta_{\mathscr{M}}\left(\alpha-\frac{1}{2}\right)=\frac{1}{\alpha}\textrm{Res}\,\zeta_{\mathscr{M}}\left(-\frac{1}{2}\right)+\textrm{FP}\,\zeta_{\mathscr{M}}\left(-\frac{1}{2}\right)+O(\alpha)\;.
\end{equation}

The Casimir force is then obtained from the energy (\ref{0a})
by differentiating with respect to the position $a$ of the piston, namely
\begin{equation}\label{10a}
F_{\textrm{Cas}}(a)=-\frac{\partial}{\partial a}E_{\textrm{Cas}}(a)\;.
\end{equation}
By using the definition in (\ref{0a}) and by keeping in mind the presence of the pole at $s=-1/2$, as shown in (\ref{12}), we obtain \cite{fucci11}
\begin{equation}\label{10b}
F_{\textrm{Cas}}(a)=-\frac{1}{2}\frac{\partial}{\partial a}\textrm{FP}\zeta_{\mathscr{M}}\left(-\frac{1}{2},a\right)-\frac{1}{2}\left(\frac{1}{\alpha}+\ln\mu^{2}\right)\frac{\partial}{\partial a}\textrm{Res}\,\zeta_{\mathscr{M}}\left(-\frac{1}{2},a\right)+O(\alpha)\;.
\end{equation}
From the last expression it is clear that the resulting force is ambiguous unless the residue of $\zeta_{\mathscr{M}}$ at $s=-1/2$ is independent of $a$.
This is a purely geometric statement since $\textrm{Res}\,\zeta_{\mathscr{M}}\left(-1/2\right)$ is proportional to the heat kernel coefficient $a_{D/2}$
of the Laplace operator on the generalized cone. It is then clear that an unambiguous evaluation of the Casimir force can be performed
when one considers an odd dimensional manifold $\mathscr{N}$ without boundary.

In order to explicitly compute the spectral $\zeta$-functions in the two regions we need to impose specific boundary conditions,
which, in turn, will provide implicit equations for the eigenvalues. In this paper we will study the case in which the scalar field is
subjected to hybrid boundary conditions.

\section{Hybrid Boundary Conditions}

In this section we will focus our attention on hybrid boundary conditions. We study the case in which the
boundary conditions on the base manifold at $r=b$ and on the piston, positioned at $r=a$, are different.
For the analysis of hybrid boundary conditions we need to distinguish between two cases.
In the first case, which we will denote as hybrid boundary conditions of first type, one imposes Dirichlet boundary conditions on the piston at $r=a$ and Neuman boundary
conditions on the base at $r=b$. In the second case, which will be denoted as hybrid boundary conditions of second type, one imposes Neuman boundary conditions on the
piston at $r=a$ and Dirichlet boundary conditions on the base at $r=b$.

It is worth pointing out that for a standard Casimir piston, for which the two chambers have the same geometry,
the two types of boundary conditions described above lead to the same results for the Casimir energy and force. There is, therefore, no need, in standard situations, to make this kind of distinction.
The case of the conical piston, however, is more involved since the two chambers do not share the same type of geometry. In this situation it is necessary
to distinguish between the two types of hybrid boundary conditions as they will lead to different results. Because of the abovementioned reasons, we will analyze, in the next sections, the
two cases separately.

\subsection{Hybrid Boundary Conditions of First Type}

In this case we impose Dirichlet boundary conditions on the piston. Therefore, in region $I$ we obtain the following
implicit equation for the eigenvalues
\begin{equation}\label{13}
J_{\nu}(\gamma_{I}a)=0\;,
\end{equation}
while in region $II$ we impose Neuman boundary conditions at $r=b$ to find
\begin{equation}\label{14}
\left\{ \begin{array}{l}
AJ_{\nu}(\gamma_{II}a)+BY_{\nu}(\gamma_{II}a)=0\\
A\left[\beta J_{\nu}(\gamma_{II}b)+\gamma_{II}bJ'_{\nu}(\gamma_{II}b)\right]+B\left[\beta Y_{\nu}(\gamma_{II}b)+\gamma_{II}bY'_{\nu}(\gamma_{II}b)\right]=0\;,\\
\end{array} \right.
\end{equation}
which has a non-trivial solution if
\begin{equation}\label{15}
\Delta_{\nu}(a,b,\gamma_{II})=J_{\nu}(\gamma_{II}a)\left[\beta Y_{\nu}(\gamma_{II}b)+\gamma_{II}bY'_{\nu}(\gamma_{II}b)\right]-Y_{\nu}(\gamma_{II}a)\left[\beta J_{\nu}(\gamma_{II}b)+\gamma_{II}bJ'_{\nu}(\gamma_{II}b)\right]=0\;,
\end{equation}
where we have defined, for typographical convenience, $\beta=(1-d)/2$.

In order to express the spectral $\zeta$-functions in the two regions, we utilize the following integral representation valid for $\Re(s)>(d+1)/2$ \cite{bordag96,bordag96a,bordag96b,esposito97,kirsten01}
\begin{equation}\label{16}
\zeta^{\mathcal{H}_{1}}_{I}(s,a)=\sum_\nu d(\nu)\frac{1}{2\pi i}\int_{\Gamma}d\kappa\left[\kappa^{2}+m^{2}\right]^{-s}\frac{\partial}{\partial \kappa}\ln \left[\kappa^{-\nu} J_{\nu}(\kappa a)\right]\;,
\end{equation}
in region $I$ and
\begin{equation}\label{17}
\zeta^{\mathcal{H}_{1}}_{II}(s,a,b)=\sum_\nu d(\nu)\frac{1}{2\pi i}\int_{\Gamma'}d\kappa\left[\kappa^{2}+m^{2}\right]^{-s}\frac{\partial}{\partial \kappa}\ln\left[\Delta_{\nu}( a, b,\kappa)\right]\;,
\end{equation}
in region $II$, where $\Gamma$ and $\Gamma'$ represent paths in the complex plane that encircle in the counterclockwise direction all the positive zeroes of, respectively,
$J_{\nu}$ and $\Delta_{\nu}$. The analytic continuation to the region $\Re(s)\leq(d+1)/2$ is obtained by first deforming the
contour of integration to the imaginary axis, which gives rise to the appearance of the modified Bessel functions $I_{\nu}(\kappa z)$ and $K_{\nu}(\kappa z)$, namely
\begin{equation}\label{17a}
\zeta^{\mathcal{H}_{1}}_{I}(s,a)=\sum_\nu d(\nu)\frac{\sin(\pi s)}{\pi}\int_{\frac{ma}{\nu}}^{\infty}d\kappa\left[\frac{\nu^{2}\kappa^{2}}{a^2}-m^{2}\right]^{-s}\frac{\partial}{\partial \kappa}\ln \left[\kappa^{-\nu}I_{\nu}(\nu \kappa)\right]\;,
\end{equation}
and
\begin{eqnarray}\label{17b}
  \zeta^{\mathcal{H}_{1}}_{II}(s,a,b)&=&\sum_\nu d(\nu)\frac{\sin(\pi s)}{\pi}\int_{\frac{m}{\nu}}^{\infty}d\kappa\left[\nu^{2}\kappa^{2}-m^{2}\right]^{-s}\frac{\partial}{\partial \kappa}\times\nonumber\\
  &&\ln\left\{I_{\nu}(\kappa a)\left[\beta K_{\nu}(\kappa b)+\kappa bK'_{\nu}(\kappa b)\right]-K_{\nu}(\kappa a)\left[\beta I_{\nu}(\kappa b)+\kappa bI'_{\nu}(\kappa b)\right]\right\}\;,
\end{eqnarray}
and by subtracting a suitable number of leading terms from the uniform asymptotic expansion of the integrand \cite{bordag96,bordag96a,bordag96b,kirsten01}.
It is important to mention that in situations when $\beta=-\nu$ the above representations for the $\zeta$-functions need to be slightly modified
and the eigenvalue $\nu=-\beta$ necessitates a separate treatment (see Section \ref{wil}).

For the $\zeta$-function in region $I$ we make use of the uniform asymptotic expansion of the modified Bessel functions $I_{\nu}(k)$ for
$\nu\to\infty$ and for $z=k/\nu$ fixed as \cite{olver54,erdelyi53}
\begin{equation}\label{17c}
I_{\nu}(\nu z)\sim\frac{1}{\sqrt{2\pi\nu}}\frac{e^{\nu\eta}}{(1+z^{2})^{1/4}}\left[1+\sum_{k=1}^{\infty}\frac{u_{k}(t)}{\nu^{k}}\right]\;,
\end{equation}
where the polynomials $u_{k}(t)$ are determined by the recurrence relation
\begin{equation}\label{18}
u_{k+1}(t)=\frac{1}{2}t^{2}(1-t^{2})u_{k}^{\prime}(t)+\frac{1}{8}\int_{0}^{t}d\tau(1-5\tau^{2})u_{k}(\tau)\;,
\end{equation}
with $u_{0}(t)=1$ and
\begin{equation}\label{19}
t=\frac{1}{\sqrt{1+z^{2}}}\;,\qquad \eta=\sqrt{1+z^{2}}+\ln\left[\frac{z}{1+\sqrt{1+z^{2}}}\right]\;.
\end{equation}
The above relations allow us to obtain the following expression for the spectral $\zeta$-function in region $I$ \cite{fucci11}
\begin{equation}\label{20}
\zeta^{\mathcal{H}_{1}}_{I}(s,a)=Z_{I}(s,a)+\sum_{i=-1}^{D}A^{\mathcal{H}_{1},\,I}_{i}(s,a)\;,
\end{equation}
where $Z_{I}(s,a)$ is an analytic function for $\Re(s)>-1$ defined as
\begin{equation}\label{21}
  Z_{I}(s,a)=\sum_{\nu}d(\nu)Z^{\nu}_{I}(s,a)\;,
\end{equation}
with
\begin{eqnarray}\label{22}
Z_{I}^{\nu}(s,a)&=&a^{2s}\nu^{-2s}\frac{\sin(\pi s)}{\pi}\int_{0}^{\infty}d\kappa\,\kappa^{-2s}\frac{\partial}{\partial \kappa}
\Bigg\{\ln\left[\kappa^{-\nu}I_{\nu}(\kappa\nu)\right]
-\ln\left[\frac{\kappa^{-\nu}}{\sqrt{2\pi\nu}}\frac{e^{\nu\eta}}{(1+\kappa^{2})^{1/4}}\right]
-\sum_{n=1}^{D}\frac{D_{n}(t)}{\nu^{n}}\Bigg\}\;,\;\;\;\;\;\;
\end{eqnarray}
and, in addition, the functions $A^{\mathcal{H}_{1},\,I}_{i}(s,a)$ can be found to be \cite{bordag96,bordag96a,fucci10,fucci11,kirsten01}
\begin{eqnarray}\label{23}
A^{\mathcal{H}_{1},\,I}_{-1}(s,a)&=&\frac{a^{2s}}{4\sqrt{\pi}}\,\frac{\Gamma\left(s-\frac{1}{2}\right)}{\Gamma(s+1)}\zeta_{\mathscr{N}}\left(s-\frac{1}{2}\right)\;,\\
A^{\mathcal{H}_{1},\,I}_{0}(s,a)&=&-\frac{a^{2s}}{4}\zeta_{\mathscr{N}}\left(s\right)\;,\\
A^{\mathcal{H}_{1},\,I}_{i}(s,a)&=&-\frac{a^{2s}}{\Gamma(s)}\zeta_{\mathscr{N}}\left(s+\frac{i}{2}\right)\sum_{b=0}^{i}x_{i,b}\frac{\Gamma\left(s+b+\frac{i}{2}\right)}{\Gamma\left(b+\frac{i}{2}\right)}\;,\label{21a}
\end{eqnarray}
where, in the previous formulas, the limit $m\to 0$ has already been taken. The polynomials $D_{n}(t)$, appearing in (\ref{22}), are defined through the expansion \cite{bordag96,bordag96a,bordag96b,fucci10,kirsten01}
\begin{equation}\label{24}
\ln\left[1+\sum_{k=1}^{\infty}\frac{u_{k}(t)}{\nu^{k}}\right]\sim\sum_{n=1}^{\infty}\frac{D_{n}(t)}{\nu^{n}}\;,
\end{equation}
and have the structure
\begin{equation}\label{25}
D_{n}(t)=\sum_{i=0}^{n}x_{i,n}t^{n+2i}\;.
\end{equation}

By following an analogous procedure of analytic continuation, the spectral $\zeta$-function in region $II$ can be written as a combination of three terms.
In fact, by using the uniform asymptotic expansion of $K_{\nu}(\nu z)$ \cite{olver54,erdelyi53}
\begin{equation}\label{26}
K_{\nu}(\nu z)\sim\sqrt{\frac{\pi}{2\nu}}\frac{e^{-\nu\eta}}{(1+z^{2})^{1/4}}\left[1+\sum_{k=1}^{\infty}(-1)^{k}\frac{u_{k}(t)}{\nu^{k}}\right]\;,
\end{equation}
and in addition
\begin{equation}\label{27}
\ln\left[\beta I_{\nu}(\nu z)+z \nu I'_{\nu}(\nu z)\right]\sim\ln\left[\sqrt{\frac{\nu}{2\pi}}e^{\nu\eta}(1+z^{2})^{1/4}\right]+\sum_{n=1}^{\infty}\frac{M_{n}(t,\beta)}{\nu^{n}}\;,
\end{equation}
with $M_{n}(t,\beta)$
\begin{equation}
  M_{n}(t,\beta)=\sum_{i=0}^{n}z_{i,n}(\beta)t^{n+2i}\;,
\end{equation}
defined by the cumulant expansion \cite{bordag96,bordag96a,bordag96b,kirsten01}
\begin{equation}\label{28}
\ln\left[1+\sum_{k=1}^{\infty}\frac{v_{k}(t)}{\nu^{k}}+\frac{\beta}{\nu}t\left(1+\sum_{k=1}^{\infty}\frac{u_{k}(t)}{\nu^{k}}\right)\right]\sim\sum_{n=1}^{\infty}\frac{M_{n}(t,\beta)}{\nu^{n}}\; ,
\end{equation}
we can write that
\begin{equation}\label{29}
  \zeta^{\mathcal{H}_{1}}_{II}(s,a,b)=Z_{II}(s,a,b)+\mathscr{F}_{\mathcal{H}_{1}}(s,a,b)+\sum_{i=-1}^{D}A_{i}^{\mathcal{H}_{1}, II}(s,a,b)\;.
\end{equation}
The polynomials $D_{n}(t)$ and $M_{n}(t,\beta)$ can be easily computed with the help of a simple computer program; some of the first few are listed
in, e.g., \cite{fucci10,fucci11}.

The function $Z_{II}$ is analytic, by construction, for $\Re(s)>-1$ and has the expression
\begin{equation}\label{30}
  Z_{II}(s,a,b)=\sum_{\nu}d(\nu)Z_{II}^{\nu}(s,a,b)\;,
\end{equation}
where $Z_{II}^{\nu}(s,a,b)$ is defined as
\begin{eqnarray}\label{31}
Z_{II}^{\nu}(s,a,b)=a^{2s}\nu^{-2s}\frac{\sin(\pi s)}{\pi}\int_{0}^{\infty}d\kappa \kappa^{-2s}\frac{\partial}{\partial\kappa}\Bigg\{\ln[K_{\nu}(\kappa\nu)]-\ln\left[\sqrt{\frac{\pi}{2\nu}}\frac{e^{-\nu\eta}}{(1+\kappa^{2})^{\frac{1}{4}}}\right]
-\sum_{n=1}^{D}(-1)^{n}\frac{D_{n}(t)}{\nu^{n}}\Bigg\}\nonumber\\
+b^{2s}\nu^{-2s}\frac{\sin(\pi s)}{\pi}\int_{0}^{\infty}d\kappa \kappa^{-2s}\frac{\partial}{\partial\kappa}\Bigg\{\ln\left[\beta I_{\nu}(\kappa\nu)+\kappa\nu I'_{\nu}(\kappa\nu)\right]-\ln\left[\sqrt{\frac{\nu}{2\pi}}e^{\nu\eta}(1+\kappa^{2})^{\frac{1}{4}}\right]-\sum_{n=1}^{D}\frac{M_{n}(t,\beta)}{\nu^{n}}\Bigg\}\;.
\end{eqnarray}
The function $\mathscr{F}_{\mathcal{H}_{1}}(s,a,b)$, instead, is defined, in the massless limit, as
\begin{equation}\label{32}
\mathscr{F}_{\mathcal{H}_{1}}(s,a,b)=\sum_\nu d(\nu)\nu^{-2s}\frac{\sin(\pi s)}{\pi}\int_{0}^{\infty}d\kappa\,\kappa^{-2s}\frac{\partial}{\partial \kappa}\ln
\left\{1-\frac{I_{\nu}(\kappa\nu a)\left[\beta K_{\nu}(\kappa\nu b)+\kappa\nu b K'_{\nu}(\kappa\nu b)\right]}{K_{\nu}(\nu\kappa a)\left[\beta I_{\nu}(\kappa\nu b)+\kappa\nu b I'_{\nu}(\kappa\nu b)\right]}\right\}\;.
\end{equation}
The domain in which the above function is analytic can be found by considering the limits $\kappa\to 0$ and $\kappa\to \infty$ \cite{fucci11,teo10}.
In the first case, namely as $\kappa\to 0$,  the integral (\ref{32}) is convergent for $\Re(s)<1/2$. As $\kappa\to \infty$
we find the following exponential behavior
\begin{equation}\label{33}
\frac{I_{\nu}(\kappa\nu a)\left[\beta K_{\nu}(\kappa\nu b)+\kappa\nu b K'_{\nu}(\kappa\nu b)\right]}{K_{\nu}(\nu\kappa a)\left[\beta I_{\nu}(\kappa\nu b)+\kappa\nu b I'_{\nu}(\kappa\nu b)\right]}\sim \exp\{-2\nu\left[\eta(b \kappa)-\eta(a \kappa)\right]\}\;,
\end{equation}
which can be obtained by exploiting the uniform asymptotic expansion of the modified Bessel functions. It is then clear
that (\ref{32}) is analytic for $\Re(s)<1/2$ since $b>a$.
Finally, the functions $A_{i}^{\mathcal{H}_{1}, II}(s,a,b)$ appearing in (\ref{26}) are expressed in terms of $A^{\mathcal{H}_{1},\,I}_{i}(s,a)$ in (\ref{23})-(\ref{21a}) as follows
\begin{eqnarray}\label{34}
  A_{i}^{\mathcal{H}_{1}, II}(s,a,b)&=&(-1)^{i}A_{i}^{\mathcal{H}_{1},\,I}(s,a)+\bar{A}_{i}^{\mathcal{H}_{1},\,I}(s,b)\;,
  \end{eqnarray}
where the functions $\bar{A}_{i}^{\mathcal{H}_{1},\,I}$ are related to the $A_{i}^{\mathcal{H}_{1},\,I}$ by the formulas
\begin{equation}\label{35}
  \bar{A}_{-1}^{\mathcal{H}_{1},\,I}(s,b)=A_{-1}^{\mathcal{H}_{1},\,I}(s,b)\;,\qquad \bar{A}_{0}^{\mathcal{H}_{1},\,I}(s,b)=-A_{0}^{\mathcal{H}_{1},\,I}(s,b)\;,
\end{equation}
and for $i\geq 1$, once the coefficients $x_{i,n}$ are replaced with $z_{i,n}(\beta)$ \cite{bordag96,bordag96a,fucci11,kirsten01},
\begin{equation}\label{36}
  \bar{A}_{i}^{\mathcal{H}_{1},\,I}(s,b)=A_{i}^{\mathcal{H}_{1},\,I}(s,b)\;.
\end{equation}

At this point, we are able to write an explicit expression for the $\zeta$-function of the conical piston endowed with hybrid
boundary conditions of the first kind in the neighborhood of $s=-1/2$. By recalling the definition (\ref{10}) and by exploiting the results (\ref{20}) and (\ref{29}) we obtain
\begin{eqnarray}\label{37}
  \zeta_{\mathscr{M}}^{\mathcal{H}_{1}}\left(\alpha-\frac{1}{2},a,b\right)&=&Z_{I}\left(-\frac{1}{2},a\right)+Z_{II}\left(-\frac{1}{2},a,b\right)+\mathscr{F}_{\mathcal{H}_{1}}\left(-\frac{1}{2},a,b\right)\nonumber\\
  &+&2\sum_{i=0}^{\left[\frac{D}{2}\right]}A_{2i}^{\mathcal{H}_{1},\,I}\left(\alpha-\frac{1}{2},a\right)+\sum_{i=-1}^{D}\bar{A}_{i}^{\mathcal{H}_{1},\,I}\left(\alpha-\frac{1}{2},b\right)\;,
\end{eqnarray}
where $[x]$ represents the integer part of $x$. The limit $\alpha\to 0$, which needs to be considered for the evaluation of the Casimir energy,
reveals the meromorphic structure of $\zeta_{\mathscr{M}}^{\mathcal{H}_{1}}$ at $s=-1/2$. From the general theory of spectral $\zeta$-functions
\cite{gilkey95,kirsten01} we have the following expansion in terms of the variable $\alpha$, as $\alpha\to 0$,
\begin{equation}\label{38}
\zeta_{\mathscr{N}}(\alpha-n)=\zeta_{\mathscr{N}}(-n)+\alpha\zeta_{\mathscr{N}}'(- n)+O(\alpha^{2})\;,\quad\textrm{for}\quad n=0,1\;\;,
\end{equation}
\begin{equation}\label{39}
\zeta_{\mathscr{N}}\left(\alpha-\frac{1}{2}\right)=\frac{1}{\alpha}\textrm{Res}\,\zeta_{\mathscr{N}}\left(-\frac{1}{2}\right)+\textrm{FP}\,\zeta_{\mathscr{N}}\left(-\frac{1}{2}\right)+O(\alpha)\;,
\end{equation}
and, for all $d+1\geq i\geq 2$,
\begin{equation}\label{40}
\zeta_{\mathscr{N}}\left(\alpha+\frac{i-1}{2}\right)=\frac{1}{\alpha}\textrm{Res}\,\zeta_{\mathscr{N}}\left(\frac{i-1}{2}\right)+\textrm{FP}\,\zeta_{\mathscr{N}}\left(\frac{i-1}{2}\right)+O(\alpha)\;.
\end{equation}
By utilizing the pole structure of $\zeta_{\mathscr{N}}$ displayed in the last formulas we obtain an expression for
the residue of $\zeta_{\mathscr{M}}^{\mathcal{H}_{1}}$ at $s=-1/2$,
\begin{eqnarray}\label{41}
  \textrm{Res}\,\zeta_{\mathscr{M}}^{\mathcal{H}_{1}}\left(-\frac{1}{2},a,b\right)&=&\frac{1}{2}\left(\frac{1}{2b}-\frac{1}{a}\right)\textrm{Res}\,\zeta_{\mathscr{N}}\left(-\frac{1}{2}\right)
  -\frac{1}{4\pi b}\zeta_{\mathscr{N}}(-1)-\frac{1}{2\pi b}\left(\frac{3}{8}-\beta\right)\zeta_{\mathscr{N}}(0)\nonumber\\
  &+&\frac{1}{a\sqrt{\pi}}\sum_{i=1}^{[D/2]}\omega_{2i}\textrm{Res}\,\zeta_{\mathscr{N}}\left(\frac{2i-1}{2}\right)+\frac{1}{2b\sqrt{\pi}}\sum_{i=2}^{D}\tilde{\omega}_{i}\textrm{Res}\,\zeta_{\mathscr{N}}\left(\frac{i-1}{2}\right)\;,
\end{eqnarray}
where we have defined, for convenience, the numerical coefficients
\begin{equation}\label{42}
  \omega_{i}=\sum_{p=0}^{i}x_{i,p}\frac{\Gamma\left(p+\frac{i-1}{2}\right)}{\Gamma\left(p+\frac{i}{2}\right)}\;,\qquad\textrm{and}\qquad \tilde{\omega}_{i}=\sum_{p=0}^{i}z_{i,p}(\beta)\frac{\Gamma\left(p+\frac{i-1}{2}\right)}{\Gamma\left(p+\frac{i}{2}\right)}\;.
\end{equation}
For the finite part at $s=-1/2$ we have, instead, the expression
\begin{eqnarray}\label{43}
  \textrm{FP}\,\zeta_{\mathscr{M}}^{\mathcal{H}_{1}}\left(-\frac{1}{2},a,b\right)&=&Z_{I}\left(-\frac{1}{2},a\right)+Z_{II}\left(-\frac{1}{2},a,b\right)+\mathscr{F}_{\mathcal{H}_{1}}\left(-\frac{1}{2},a,b\right)
  -\frac{1}{2a}\left[\textrm{FP}\,\zeta_{\mathscr{N}}\left(-\frac{1}{2}\right)+\ln a^{2}\textrm{Res}\,\zeta_{\mathscr{N}}\left(-\frac{1}{2}\right)\right]\nonumber\\
  &+&\frac{1}{a\sqrt{\pi}}\sum_{i=1}^{[D/2]}\Bigg[\omega_{2i}\textrm{FP}\,\zeta_{\mathscr{N}}\left(\frac{2i-1}{2}\right)+\omega_{2i}\left(\ln a^{2}+\gamma+2\ln 2-2\right)\textrm{Res}\,\zeta_{\mathscr{N}}\left(\frac{2i-1}{2}\right)\nonumber\\
  &+&\Omega_{2i}\textrm{Res}\,\zeta_{\mathscr{N}}\left(\frac{2i-1}{2}\right)\Bigg]-\frac{1}{4\pi b}\left[\left(2\ln 2+1\right)\zeta_{\mathscr{N}}(-1)+\zeta'_{\mathscr{N}}(-1)\right]+\frac{1}{4b}\textrm{FP}\,\zeta_{\mathscr{N}}\left(-\frac{1}{2}\right)\nonumber\\
  &-&\frac{1}{2\pi b}\left(\frac{3}{8}-\beta\right)\left[\left(2\ln 2-2\right)\zeta_{\mathscr{N}}(0)+\zeta'_{\mathscr{N}}(0)\right]-\frac{7}{24\pi b}\zeta_{\mathscr{N}}(0)\nonumber\\
  &+&\frac{1}{2b\sqrt{\pi}}\sum_{i=2}^{D}\Bigg[\tilde{\omega}_{i}\textrm{FP}\,\zeta_{\mathscr{N}}\left(\frac{i-1}{2}\right)+\tilde{\omega}_{i}\left(\ln b^{2}+\gamma+2\ln 2-2\right)\textrm{Res}\,\zeta_{\mathscr{N}}\left(\frac{i-1}{2}\right)\nonumber\\
  &+&\tilde{\Omega}_{i}\textrm{Res}\,\zeta_{\mathscr{N}}\left(\frac{i-1}{2}\right)\Bigg]\;,
\end{eqnarray}
where $\gamma$ represents the Euler-Mascheroni constant, and we have introduced the constants
\begin{equation}\label{44}
  \Omega_{i}=\sum_{p=0}^{i}x_{i,p}\frac{\Gamma\left(p+\frac{i-1}{2}\right)}{\Gamma\left(p+\frac{i}{2}\right)}\Psi\left(p+\frac{i-1}{2}\right)\;,\qquad\textrm{and}\qquad
  \tilde{\Omega}_{i}=\sum_{p=0}^{i}z_{i,p}(\beta)\frac{\Gamma\left(p+\frac{i-1}{2}\right)}{\Gamma\left(p+\frac{i}{2}\right)}\Psi\left(p+\frac{i-1}{2}\right)\;,
\end{equation}
with $\Psi(x)$ representing the logarithmic derivative of the Euler gamma function.

The definition (\ref{10b}), together with the results (\ref{41}) and (\ref{43}), allows us to express the Casimir force on the piston
when hybrid boundary conditions of first type are imposed. More explicitly we have
\begin{eqnarray}\label{45}
  F_{\textrm{Cas}}^{\mathcal{H}_{1}}(a,b)&=&-\frac{1}{2}Z'_{I}\left(-\frac{1}{2},a\right)-\frac{1}{2}Z'_{II}\left(-\frac{1}{2},a,b\right)-\frac{1}{2}\mathscr{F}'_{\mathcal{H}_{1}}\left(-\frac{1}{2},a,b\right)
  +\frac{1}{4a^{2}}\Bigg[\left(2-\ln a^{2}\right)\textrm{Res}\,\zeta_{\mathscr{N}}\left(-\frac{1}{2}\right)\nonumber\\
  &-&\textrm{FP}\,\zeta_{\mathscr{N}}\left(-\frac{1}{2}\right)\Bigg]
  +\frac{1}{2a^{2}\sqrt{\pi}}\sum_{i=1}^{[D/2]}\Bigg[\omega_{2i}\textrm{FP}\,\zeta_{\mathscr{N}}\left(\frac{2i-1}{2}\right)-\omega_{2i}\left(4-\ln a^{2}-\gamma-2\ln 2\right)\textrm{Res}\,\zeta_{\mathscr{N}}\left(\frac{2i-1}{2}\right)\nonumber\\
  &+&\Omega_{2i}\textrm{Res}\,\zeta_{\mathscr{N}}\left(\frac{2i-1}{2}\right)\Bigg]-\frac{1}{4a^{2}}\left(\frac{1}{\alpha}+\ln\mu^{2}\right)\left[\textrm{FP}\,\zeta_{\mathscr{N}}\left(-\frac{1}{2}\right)
  -\frac{2}{\sqrt{\pi}}\sum_{i=1}^{[D/2]}\omega_{2i}\textrm{Res}\,\zeta_{\mathscr{N}}\left(\frac{2i-1}{2}\right)\right]+O(\alpha)\;.\nonumber\\
\end{eqnarray}
Here the prime denotes differentiation with respect to the variable $a$. Let us point out that the term in the last expression, proportional to
$(1/\alpha+\ln\mu^{2})$, represents the ambiguity that appears in general when computing the Casimir force.

\subsection{Hybrid Boundary Conditions of Second Type}

In order to consider this case we impose Neuman boundary conditions on the piston, which leads, in region $I$, to the following
 condition
\begin{equation}\label{46}
\beta J_{\nu}(\gamma_{I}a)+a\gamma_{I}J'_{\nu}(\gamma_{I}a)=0\;.
\end{equation}
In region $II$ we impose Dirichlet boundary conditions at $r=b$ and we obtain
\begin{equation}\label{47}
\left\{ \begin{array}{l}
A\left[\beta J_{\nu}(\gamma_{I}a)+a\gamma_{I}J'_{\nu}(\gamma_{I}a)\right]+B\left[\beta Y_{\nu}(\gamma_{I}a)+a\gamma_{I}Y'_{\nu}(\gamma_{I}a)\right]=0\\
AJ_{\nu}(\gamma_{II}b)+BY'_{\nu}(\gamma_{II}b)=0\;,\\
\end{array} \right.
\end{equation}
which provides an implicit equation for the eigenvalues
\begin{equation}\label{48}
Y_{\nu}(\gamma_{II}b)\left[\beta J_{\nu}(\gamma_{II}a)+a\gamma_{II}J'_{\nu}(\gamma_{II}a)\right]-J_{\nu}(\gamma_{II}b)\left[\beta Y_{\nu}(\gamma_{II}a)+\gamma_{II}a Y'_{\nu}(\gamma_{II}a)\right]=0\;.
\end{equation}
The spectral $\zeta$-functions associated with region $I$ and region $II$ are expressed, in the same fashion as in the previous section, in terms of a complex integral representation
valid for $\Re(s)>(d+1)/2$ \cite{bordag96,bordag96a,bordag96b,esposito97,kirsten01}. By deformation of the contour of integration we obtain the following relations
\begin{equation}\label{48a}
\zeta^{\mathcal{H}_{2}}_{I}(s,a)=\sum_{\nu}d(\nu)\frac{\sin(\pi s)}{\pi}\int_{\frac{ma}{\nu}}^{\infty}d\kappa\left[\frac{\nu^{2}\kappa^{2}}{a^2}-m^{2}\right]^{-s}\frac{\partial}{\partial \kappa}\ln \left\{\kappa^{-\nu}\left[\beta I_{\nu}(\nu \kappa)+\nu \kappa I'_{\nu}(\nu \kappa)\right]\right\}\;,
\end{equation}
for region $I$ and
\begin{eqnarray}\label{48b}
  \zeta^{\mathcal{H}_{2}}_{II}(s,a,b)&=&\sum_{\nu}d(\nu)\frac{\sin(\pi s)}{\pi}\int_{\frac{m}{\nu}}^{\infty}d\kappa\left[\nu^{2}\kappa^{2}-m^{2}\right]^{-s}\frac{\partial}{\partial \kappa}\nonumber\\
  &&\ln\left\{K_{\nu}(\kappa b)\left[\beta I_{\nu}(\kappa a)+\kappa aI'_{\nu}(\kappa a)\right]-I_{\nu}(\kappa b)\left[\beta K_{\nu}(\kappa a)+\kappa aK'_{\nu}(\kappa a)\right]\right\}\;,
\end{eqnarray}
for region $II$. Once again, the above representations are valid for $\nu>-\beta$, and a separate treatment is needed if $\nu=-\beta$.

Since the analytic continuation to the region $\Re(s)\leq (d+1)/2$ is obtained
by following the procedure delineated for hybrid boundary conditions of first type, we will be more direct and simply present the most important results.

For the spectral $\zeta$-function in region $I$, we utilize the uniform asymptotic expansion (\ref{27}) to obtain the expression
\begin{equation}\label{49}
  \zeta_{I}^{\mathcal{H}_{2}}(s,a)=W_{I}(s,a)+\sum_{i=1}^{D}\bar{A}^{\mathcal{H}_{1},\,I}_{i}(s,a)\;,
\end{equation}
where $\bar{A}^{\mathcal{H}_{1},\,I}_{i}(s,a)$ are functions defined in (\ref{35}) and (\ref{36}), while
\begin{equation}\label{50}
  W_{I}(s,a)=\sum_{\nu}d(\nu)W_{I}^{\nu}(s,a)\;,
\end{equation}
is an analytic function for $\Re(s)> -1$, with $W_{I}^{\nu}(s,a)$ having the following integral representation
\begin{eqnarray}\label{51}
  W_{I}^{\nu}(s,a)&=&a^{2s}\nu^{-2s}\frac{\sin(\pi s)}{\pi}\int_{0}^{\infty}d\kappa \kappa^{-2s}\frac{\partial}{\partial\kappa}\Bigg\{\ln\left[\left(\beta I_{\nu}(\kappa\nu)+\kappa\nu I'_{\nu}(\kappa\nu)\right)\right]\nonumber\\
  &-&\ln\left[\sqrt{\frac{\nu}{2\pi}}e^{\nu\eta}(1+\kappa^{2})^{\frac{1}{4}}\right]
-\sum_{n=1}^{D}\frac{M_{n}(t,\beta)}{\nu^{n}}\Bigg\}\;.
\end{eqnarray}

For the spectral $\zeta$-function in region $II$ we obtain a result which is similar to (\ref{29}), namely
\begin{equation}\label{52}
  \zeta^{\mathcal{H}_{2}}_{II}(s,a,b)=W_{II}(s,a,b)+\mathscr{F}_{\mathcal{H}_{2}}(s,a,b)+\sum_{i=-1}^{D}A_{i}^{\mathcal{H}_{2}, II}(s,a,b)\;,
\end{equation}
which can be obtained by making use of the expansion (\ref{17c}) and the following \cite{gradshtein07,olver54}
\begin{equation}\label{53}
\ln\left[-\beta K_{\nu}(\nu\kappa)-\nu\kappa K'_{\nu}(\nu\kappa)\right]\sim\ln\left[\sqrt{\frac{\pi\nu}{2}}e^{-\nu\eta}(1+\kappa^{2})^{1/4}\right]+\sum_{n=1}^{\infty}(-1)^{n}\frac{M_{n}(t,\beta)}{\nu^{n}}\;.
\end{equation}
The function $W_{II}(s,a,b)$, which is analytic for $\Re(s)>-1$, is written as
\begin{equation}\label{54}
  W_{II}(s,a,b)=\sum_{\nu}d(\nu)W_{II}^{\nu}(s,a,b)\;,
\end{equation}
where
\begin{eqnarray}\label{55}
  W_{II}^{\nu}(s,a,b)&=&a^{2s}\nu^{-2s}\frac{\sin(\pi s)}{\pi}\int_{0}^{\infty}d\kappa \kappa^{-2s}\frac{\partial}{\partial\kappa}\Bigg\{\ln\left[-\beta K_{\nu}(\kappa\nu)-\kappa\nu K'_{\nu}(\kappa\nu)\right]-\ln\left[\sqrt{\frac{\pi\nu}{2}}e^{-\nu\eta}(1+\kappa^{2})^{\frac{1}{4}}\right]\nonumber\\
  &-&\sum_{n=1}^{D}(-1)^{n}\frac{M_{n}(t,\beta)}{\nu^{n}}\Bigg\}+b^{2s}\nu^{-2s}\frac{\sin(\pi s)}{\pi}\int_{0}^{\infty}d\kappa \kappa^{-2s}\frac{\partial}{\partial\kappa}\Bigg\{\ln\left[\kappa^{-\nu}I_{\nu}(\kappa\nu)\right]\nonumber\\
  &-&\ln\left[\frac{\kappa^{-\nu}}{\sqrt{2\pi\nu}}\frac{e^{\nu\eta}}{(1+\kappa^{2})^{\frac{1}{4}}}\right]
  -\sum_{n=1}^{D}\frac{D_{n}(t)}{\nu^{n}}\Bigg\}\;.
\end{eqnarray}
The function $\mathscr{F}_{\mathcal{H}_{2}}(s,a,b)$ that appears in (\ref{52}) has the form
\begin{eqnarray}\label{56}
  \mathscr{F}_{\mathcal{H}_{2}}(s,a,b)=\sum_{\nu}d(\nu)\nu^{-2s}\frac{\sin(\pi s)}{\pi}\int_{0}^{\infty}d\kappa \kappa^{-2s}\frac{\partial}{\partial\kappa}
  \ln\left\{1-\frac{K_{\nu}( \kappa\nu b)\left[\beta I_{\nu}(\kappa\nu a)+\kappa\nu a I'_{\nu}(\kappa\nu a)\right]}{I_{\nu}(\kappa\nu b)\left[\beta K_{\nu}(\kappa\nu a)+\kappa\nu a K'_{\nu}(\kappa\nu a)\right]}\right\}\;.
\end{eqnarray}
By an argument similar to the one used to determine the domain of analyticity of $\mathscr{F}_{\mathcal{H}_{1}}(s,a,b)$ in (\ref{32}),
one can show that $\mathscr{F}_{\mathcal{H}_{2}}(s,a,b)$ constitutes an analytic function in the region $\Re(s)<1/2$. In fact, the behavior of the integrand as $\kappa\to 0$ leads to the condition $\Re(s)<1/2$.
As $\kappa\to\infty$, instead, the integrand tends to zero exponentially fast, which does not impose any condition on the range of $s$.
This means, in particular, that $s=-1/2$ can be substituted in (\ref{56}).
The terms $A_{i}^{\mathcal{H}_{2}, II}(s,a,b)$ are found to be related to $\bar{A}^{\mathcal{H}_{1},\,I}_{i}$ and $A^{\mathcal{H}_{1},\,I}_{i}$ as follows
\begin{equation}
  A_{i}^{\mathcal{H}_{2}, II}(s,a,b)=(-1)^{i}\bar{A}^{\mathcal{H}_{1},\,I}_{i}(s,a)+A^{\mathcal{H}_{1},\,I}_{i}(s,b)\;.
\end{equation}

By combining the results obtained in (\ref{49}) and (\ref{52}) we can write the spectral $\zeta$-function for the conical piston associated to
hybrid boundary conditions of second type as
\begin{eqnarray}\label{57}
  \zeta_{\mathscr{M}}^{\mathcal{H}_{2}}\left(\alpha-\frac{1}{2},a,b\right)&=&W_{I}\left(-\frac{1}{2},a\right)+W_{II}\left(-\frac{1}{2},a,b\right)+\mathscr{F}_{\mathcal{H}_{2}}\left(-\frac{1}{2},a,b\right)\nonumber\\
  &+&2\sum_{i=0}^{\left[\frac{D}{2}\right]}\bar{A}_{2i}^{\mathcal{H}_{1},\,I}\left(\alpha-\frac{1}{2},a\right)+\sum_{i=-1}^{D}A_{i}^{\mathcal{H}_{1},\,I}\left(\alpha-\frac{1}{2},b\right)\;.
\end{eqnarray}
By using the expansions (\ref{38})-(\ref{40}) in the previous expression, one can extract the residue and finite part of $\zeta_{\mathscr{M}}^{\mathcal{H}_{2}}(s)$
at the point $s=-1/2$. More explicitly we obtain, for the residue,
\begin{eqnarray}\label{58}
  \textrm{Res}\,\zeta_{\mathscr{M}}^{\mathcal{H}_{2}}\left(-\frac{1}{2},a,b\right)&=&\frac{1}{2}\left(\frac{1}{a}-\frac{1}{2b}\right)\textrm{Res}\,\zeta_{\mathscr{N}}\left(-\frac{1}{2}\right)
  -\frac{1}{4\pi b}\zeta_{\mathscr{N}}(-1)+\frac{1}{16\pi b}\zeta_{\mathscr{N}}(0)\nonumber\\
  &+&\frac{1}{a\sqrt{\pi}}\sum_{i=1}^{[D/2]}\tilde{\omega}_{2i}\textrm{Res}\,\zeta_{\mathscr{N}}\left(\frac{2i-1}{2}\right)
  +\frac{1}{2b\sqrt{\pi}}\sum_{i=2}^{D}\omega_{i}\textrm{Res}\,\zeta_{\mathscr{N}}\left(\frac{i-1}{2}\right)\;,
\end{eqnarray}
and, for the finite part,
\begin{eqnarray}\label{59}
  \textrm{FP}\,\zeta_{\mathscr{M}}^{\mathcal{H}_{2}}\left(-\frac{1}{2},a,b\right)&=&W_{I}\left(-\frac{1}{2},a\right)+W_{II}\left(-\frac{1}{2},a,b\right)+\mathscr{F}_{\mathcal{H}_{2}}\left(-\frac{1}{2},a,b\right)
  +\frac{1}{2a}\left[\textrm{FP}\,\zeta_{\mathscr{N}}\left(-\frac{1}{2}\right)+\ln a^{2} \textrm{Res}\,\zeta_{\mathscr{N}}\left(-\frac{1}{2}\right)\right]\nonumber\\
  &+&\frac{1}{a\sqrt{\pi}}\sum_{i=1}^{[D/2]}\Bigg[\tilde{\omega}_{2i}\textrm{FP}\,\zeta_{\mathscr{N}}\left(\frac{2i-1}{2}\right)+\tilde{\omega}_{2i}\left(\ln a^{2}+\gamma+2\ln 2-2\right)\textrm{Res}\,\zeta_{\mathscr{N}}\left(\frac{2i-1}{2}\right)\nonumber\\
  &+&\tilde{\Omega}_{2i}\textrm{Res}\,\zeta_{\mathscr{N}}\left(\frac{2i-1}{2}\right)\Bigg]-\frac{1}{4\pi b}\left[(2\ln 2+1)\zeta_{\mathscr{N}}(-1)+\zeta'_{\mathscr{N}}(-1)\right]
  -\frac{1}{4b}\textrm{FP}\,\zeta_{\mathscr{N}}\left(-\frac{1}{2}\right)\nonumber\\
  &+&\frac{1}{16\pi b}\left[\zeta'_{\mathscr{N}}(0)+\left(2\ln 2-\frac{16}{3}\right)\zeta_{\mathscr{N}}(0)\right]+
  \frac{1}{2b\sqrt{\pi}}\sum_{i=2}^{D}\Bigg[\omega_{i}\textrm{FP}\,\zeta_{\mathscr{N}}\left(\frac{i-1}{2}\right)\nonumber\\
  &+&\omega_{i}\left(\ln b^{2}+\gamma+2\ln 2-2\right)\textrm{Res}\,\zeta_{\mathscr{N}}\left(\frac{i-1}{2}\right)
  +\Omega_{i}\textrm{Res}\,\zeta_{\mathscr{N}}\left(\frac{i-1}{2}\right)\Bigg]\;.
\end{eqnarray}

Thanks to the obtained expressions for the residue and finite part of $\zeta_{\mathscr{M}}^{\mathcal{H}_{2}}$ at $s=-1/2$, we find, by using (\ref{10b}),
the Casimir force of the piston for hybrid boundary conditions of second type, namely
\begin{eqnarray}\label{60}
  F_{\textrm{Cas}}^{\mathcal{H}_{2}}(a,b)&=&-\frac{1}{2}W'_{I}\left(-\frac{1}{2},a\right)-\frac{1}{2}W'_{II}\left(-\frac{1}{2},a,b\right)-\frac{1}{2}\mathscr{F}'_{\mathcal{H}_{2}}\left(-\frac{1}{2},a,b\right)
  -\frac{1}{4a^{2}}\Bigg[\left(2-\ln a^{2}\right)\textrm{Res}\,\zeta_{\mathscr{N}}\left(-\frac{1}{2}\right)\nonumber\\
  &-&\textrm{FP}\,\zeta_{\mathscr{N}}\left(-\frac{1}{2}\right)\Bigg]+\frac{1}{2a^{2}\sqrt{\pi}}\sum_{i=1}^{[D/2]}\Bigg[\tilde{\omega}_{2i}\textrm{FP}\,\zeta_{\mathscr{N}}\left(\frac{2i-1}{2}\right)-\tilde{\omega}_{2i}\left(4-\ln a^{2}-\gamma-2\ln 2\right)\textrm{Res}\,\zeta_{\mathscr{N}}\left(\frac{2i-1}{2}\right)\nonumber\\
  &+&\tilde{\Omega}_{2i}\textrm{Res}\,\zeta_{\mathscr{N}}\left(\frac{2i-1}{2}\right)\Bigg]+\frac{1}{4a^{2}}\left(\frac{1}{\alpha}+\ln\mu^{2}\right)\left[\textrm{FP}\,\zeta_{\mathscr{N}}\left(-\frac{1}{2}\right)
  -\frac{2}{\sqrt{\pi}}\sum_{i=1}^{[D/2]}\tilde{\omega}_{2i}\textrm{Res}\,\zeta_{\mathscr{N}}\left(\frac{2i-1}{2}\right)\right]+O(\alpha)\;.\nonumber\\
\end{eqnarray}

The results (\ref{45}) and (\ref{60}) represent the Casimir force on the piston when one imposes, respectively, hybrid boundary conditions of first and second type.
These formulas are valid for any dimension $D$ and for any smooth, compact manifold $\mathscr{N}$ with or without boundary.
Let us point out that (\ref{45}) and (\ref{60}) are very general and, thus, given in terms of the spectral $\zeta$-function on $\mathscr{N}$ which is intimately related to the geometry of the piston.
It is clear that more explicit
results can only be obtained once the piston $\mathscr{N}$ has been specified.
We would like to stress, at this point, that the results for the Casimir force for the two types of hybrid boundary conditions are essentially different.
This is in contrast to the situation that one encounters when dealing with standard Casimir pistons for which the chambers have the same geometry.
In standard pistons, in fact, there is no distinction between the two types of boundary conditions since they lead to the same Casimir force.
This non-standard behavior of the conical piston seems to be a novel feature which is due to the fact that the two chambers have different geometry.

Let us also notice that the expressions (\ref{45}) and (\ref{60}) contain explicitly the terms that are responsible for the ambiguity in the Casimir force.
These terms are proportional to the heat kernel coefficients $a_{(d+1)/2-i}$ of the manifold $\mathscr{N}$ with $0\leq i\leq [(d+1)/2]$.
It is clear that the ambiguity in the prediction of the force disappears if the manifold $\mathscr{N}$ is even-dimensional without boundary as we have mentioned earlier \cite{fucci11}.

\section{Limiting Cases}

In this section we study in detail the behavior of the Casimir force on the conical piston in two particular cases.
In the first case we consider the limit as $a\to\infty$ and $b\to\infty$, or, more precisely, when the ratio $b/a\to 1$.
In this situation the piston, positioned at $a$, approaches the manifold $\mathscr{N}$ at $b$ in such a way to approximate the configuration
of two parallel plates with hybrid boundary conditions. The second case that we investigate is the limit as $a\to 0$, which
fully describes the interaction of the piston when it approaches the conical singularity ar $r=0$. In order to be succinct,
we will describe in detail only the case of hybrid boundary conditions of first type since the other case can be treated in a similar manner.

\subsection{Large $a$ and $b$, namely $b/a\to 1$}

From the expression of the Casimir force in (\ref{45}) we observe that all the terms, including $Z'_{I}$ and $Z'_{II}$, become negligible
as $a\to \infty$ since they are proportional to $a^{-2}$. It is therefore clear that the only term that will contribute to the force, when both $a$ and $b$
are large, is the function $\mathscr{F}'_{\mathcal{H}_{1}}\left(-1/2,a,b\right)$. This means, in particular, that in this limit we can write
\begin{equation}\label{61}
  F_{\textrm{Cas}}^{\mathcal{H}_{1}}(a,b)\sim-\frac{1}{2\pi}\sum_{\nu}d(\nu)\nu\int_{0}^{\infty}d\kappa \frac{\partial}{\partial a}\ln
  \left\{1-\frac{I_{\nu}(\kappa\nu a)\left[\beta K_{\nu}(\kappa\nu b)+\kappa\nu b K'_{\nu}(\kappa\nu b)\right]}{K_{\nu}(\nu\kappa a)\left[\beta I_{\nu}(\kappa\nu b)+\kappa\nu b I'_{\nu}(\kappa\nu b)\right]}\right\}\;.
\end{equation}
The derivative with respect to the parameter $a$ inside the integral, which we will denote by $\mathcal{Q}(\kappa,a,b)$, is easily performed leading to the result
\begin{equation}\label{62}
  \mathcal{Q}(\kappa,a,b)=\frac{-\beta K_{\nu}(\kappa\nu b)-\kappa\nu b K'_{\nu}(\kappa\nu b)}{a K^{2}_{\nu}(\kappa\nu a)\left[\beta I_{\nu}(\kappa\nu b)+\kappa\nu b I'_{\nu}(\kappa\nu b)\right]}\left\{1+\frac{I_{\nu}(\kappa\nu a)\left[-\beta K_{\nu}(\kappa\nu b)-\kappa\nu b K'_{\nu}(\kappa\nu b)\right]}{K_{\nu}(\nu\kappa a)\left[\beta I_{\nu}(\kappa\nu b)+\kappa\nu b I'_{\nu}(\kappa\nu b)\right]}\right\}^{-1}\;.
\end{equation}
By performing a change of variable $\kappa\to\kappa/a$, one obtains
\begin{equation}\label{63}
  F_{\textrm{Cas}}^{\mathcal{H}_{1}}(a,b)\sim-\frac{1}{2\pi a^{2}}\sum_{\nu}d(\nu)\nu\int_{0}^{\infty}d\kappa\,\mathcal{Q}\left(\frac{\nu\kappa}{a},1,\frac{b}{a}\right)\;.
\end{equation}

In order to study the behavior of $F_{\textrm{Cas}}^{\mathcal{H}_{1}}$ when $a/b\to 1$, it is convenient to exploit the uniform asymptotic expansion
of the modified Bessel functions which provides the expression \cite{fucci11,teo10}
\begin{equation}\label{64}
  \mathcal{Q}\left(\frac{\nu\kappa}{a},1,\frac{b}{a}\right)\sim 2\nu(1+\kappa^{2})^{\frac{1}{4}}\sum_{n=1}^{\infty}e^{-2\nu n\left[\eta\left(\frac{\kappa b}{a}\right)-\eta(\kappa)\right]}
  \sum_{i=0}^{\infty}\frac{q_{i,n}(t,a,b)}{\nu^{i}}\;,
\end{equation}
where $q_{i,n}(t,a,b)$ are polynomials in $t$ and $q_{0,n}(t,a,b)=1$. By  expressing the exponential function that appears in the previous asymptotic expansion in terms of
a Mellin-Barnes integral and by recalling the definition of $\zeta_{\mathscr{N}}$, we are able to rewrite the leading contribution to the Casimir force in the form \cite{fucci11,teo10}
\begin{equation}\label{e}
F^{\textrm{Dir}}_{\textrm{Cas}}(q)\sim-\frac{1}{2\pi^{2}i\,a^{2}}\int_{c-i\infty}^{c+i\infty}d\alpha\,\Gamma(\alpha)(2)^{-\alpha}\zeta_{R}(\alpha)\zeta_{\mathscr{N}}\left(\frac{\alpha}{2}-1\right)\int_{0}^{\infty}d\kappa\sqrt{1+\kappa^{2}}\left[\eta\left((q+1)\kappa\right)-\eta(\kappa)\right]^{-\alpha}\;,
\end{equation}
which is a well defined expression for $\Re(c)>2$. Here we have introduced the new variable $q=b/a-1$ such that the limit of interest becomes $q\to 0$.
By closing the contour of integration to the right, the leading contribution to the Casimir force comes from the rightmost pole of $\zeta_\mathscr{N}$ which is at $\alpha=d+2$ \cite{fucci11}.
By noticing that $\eta\left((q+1)\kappa\right)-\eta(\kappa)=q\sqrt{1+\kappa^{2}}+O(q^{2})$, the application of the residue theorem to the integral in (\ref{e}) then leads to the result \cite{fucci11}
\begin{equation}\label{65}
  F^{\textrm{Dir}}_{\textrm{Cas}}(q)\sim-\frac{\Gamma(D+1)\zeta_{R}(D+1)}{2^{D+1}\sqrt{\pi}\,\Gamma\left(\frac{D}{2}\right)}\frac{\mathscr{A}^{\mathscr{N}}_{0}}{q^{D+1}}\;,
\end{equation}
where $\mathscr{A}^{\mathscr{N}}_{0}$ represents the zeroth order coefficient of the heat kernel asymptotic expansion of the Laplacian $\Delta_{\mathscr{N}}$.
We immediately see that the sign in the expression (\ref{65}) indicates that the piston at $a$ is repelled from the base $\mathscr{N}$ positioned at $b$ when $b/a\to 1$ . This is in complete agreement
with the well known fact that parallel plates develop a negative force when hybrid boundary conditions are imposed (see e.g. \cite{bordag09}).
Results similar to the ones in (\ref{63}) and (\ref{64}) can be obtained for the case of hybrid boundary conditions of second type. The explicit evaluation
of the force, then, follows the same ideas outlined above.

\subsection{Small $a$ Behavior}

The result (\ref{45}) shows that, as $a\to 0$, all but one term are proportional to $a^{-2}$, with the explicit behavior depending on the specific geometry of the piston $\mathscr{N}$. The exception
is the function $\mathscr{F}'_{\mathcal{H}_{1}}\left(-1/2,a,b\right)$ for which a more detailed analysis is needed. From the definition (\ref{32}) we can write that
\begin{equation}\label{66}
-\frac{1}{2}\mathscr{F}'_{\mathcal{H}_{1}}\left(-\frac{1}{2},a,b\right)=-\frac{1}{2\pi}\sum_{\nu}d(\nu)\nu\int_{0}^{\infty}d\kappa \,\mathcal{Q}\left(\frac{\kappa}{\nu},a,b\right)\;,
\end{equation}
where we have employed the change of variable $\kappa\to\kappa/\nu$. By using the small $\kappa$ expansion of the modified Bessel functions \cite{erdelyi53,gradshtein07}
one obtains an expression to the leading order in $a$
\begin{equation}\label{67}
  -\frac{1}{2}\mathscr{F}'_{\mathcal{H}_{1}}\left(-\frac{1}{2},a,b\right)\sim \frac{2}{\pi}\sum_{\nu}d(\nu)\frac{a^{2\nu-1}}{\Gamma^{2}(\nu)}\int_{0}^{\infty}d\kappa\,
  \frac{\beta K_{\nu}(\kappa\nu b)+\kappa\nu b K'_{\nu}(\kappa\nu b)}{\beta I_{\nu}(\kappa\nu b)+\kappa\nu b I'_{\nu}(\kappa\nu b)}\left(\frac{\kappa}{2}\right)^{2\nu}\;.
\end{equation}
The integral that appears in the previous formula is convergent. In fact, as $\kappa\to 0$, the integrand has the following asymptotic behavior
\begin{equation}\label{68}
  \frac{\beta K_{\nu}(\kappa\nu b)+\kappa\nu b K'_{\nu}(\kappa\nu b)}{\beta I_{\nu}(\kappa\nu b)+\kappa\nu b I'_{\nu}(\kappa\nu b)}\left(\frac{\kappa}{2}\right)^{2\nu}\sim
  \frac{1}{2}\Gamma(\nu+1)\Gamma(\nu)\left(\frac{\beta-\nu}{\beta+\nu}\right)b^{-2\nu}\;,
\end{equation}
while for $\kappa\to \infty$ we have
\begin{equation}\label{69}
  \frac{\beta K_{\nu}(\kappa\nu b)+\kappa\nu b K'_{\nu}(\kappa\nu b)}{\beta I_{\nu}(\kappa\nu b)+\kappa\nu b I'_{\nu}(\kappa\nu b)}\left(\frac{\kappa}{2}\right)^{2\nu}\sim
  \pi e^{-2\kappa}\left(\frac{\kappa}{2}\right)^{2\nu}\;,
\end{equation}
which is exponentially decaying in $\kappa$.
We can then conclude that for $\nu > -1/2$ and $\beta\neq-\nu$, which is within the assumptions of our work,
the contributions coming from $\mathscr{F}'_{\mathcal{H}_{1}}$ are subleading as $a\to 0$ \cite{fucci11}.
A similar analysis can be performed along the same lines for hybrid boundary conditions of second type, obtaining in this case
that $\mathscr{F}'_{\mathcal{H}_{2}}$ becomes subleading when $a\to 0$.

\section{The $d$-dimensional Sphere as Piston}\label{wil}

In this section we analyze the case in which the base manifold is represented by a $d$-dimensional sphere. In this particular
situation the eigenvalues of the Laplacian $\Delta_{\mathscr{N}}$ are known to be
\begin{equation}\label{61a}
\nu=\left(l+\frac{d-1}{2}\right)\;,
\end{equation}
with $l\geq 0$, and the eigenfunctions are hyperspherical harmonics with degeneracy
\begin{equation}\label{62a}
d(l)=(2l+d-1)\frac{(l+d-2)!}{l!(d-1)!}\;.
\end{equation}
The spectral $\zeta$-function on $\mathscr{N}$ can then be written as a linear combination of Hurwitz $\zeta$-functions as follows \cite{bordag96,bordag96a}
\begin{equation}\label{63a}
\zeta_{{\mathscr{N}}}(s)=2\sum_{\alpha=0}^{d-1}e_{\alpha}\zeta_{H}\left(2s-\alpha-1,\frac{d-1}{2}\right)\;,
\end{equation}
where the coefficients $e_{\alpha}$ are defined according to the relation
\begin{equation}\label{64a}
\frac{(l+d-2)!}{l!(d-1)!}=\sum_{\alpha=0}^{d-1}e_{\alpha}\left(l+\frac{d-1}{2}\right)^{\alpha}\;.
\end{equation}
The expression of $\zeta_{\mathscr{N}}$ obtained above will be used in order to specialize the general results obtained in (\ref{45}) and
(\ref{60}) to the $d$-dimensional sphere. First of all, we notice that for $s=-m/2$, with $m\geq -1$, the function $\zeta_{\mathscr{N}}$
possesses no poles, and by utilizing the equation (\ref{63a}), we obtain the result \cite{barnes03a,barnes03b,chang,dowker94,fucci11}
\begin{equation}\label{65a}
\zeta_{\mathscr{N}}\left(-\frac{m}{2}\right)=-2\sum_{\alpha=0}^{d-1}\frac{e_{\alpha}}{m+\alpha+2}B_{m+\alpha+2}\left(\frac{d-1}{2}\right)\;,
\end{equation}
where $B_{n}(q)$ are the Bernoulli polynomials \cite{gradshtein07}.

The residue and finite part of $\zeta_{\mathscr{N}}$ at the points $s=m/2$ with $m\geq 0$, can be obtained from (\ref{63a}) by recalling that the Hurwitz $\zeta$-function
has a simple pole at $s=1$. This remark allows one to show that at $s=m/2$ with $d\geq m\geq 2$, the function $\zeta_{\mathscr{N}}(m/2)$ has the residue \cite{fucci11}
\begin{equation}\label{66a}
  \textrm{Res}\,\zeta_{\mathscr{N}}\left(\frac{m}{2}\right)=e_{m-2}\;.
\end{equation}
The finite part of $\zeta_{\mathscr{N}}(m/2)$ has, instead, the form \cite{fucci11}
\begin{equation}\label{67a}
\textrm{FP}\,\zeta_{\mathscr{N}}\left(\frac{m}{2}\right)=2\sum_{{\alpha=0 \atop \alpha\neq m-2}}^{d-1}e_{\alpha}\zeta_{H}\left(m-\alpha-1,\frac{d-1}{2}\right)+2e_{m-2}\left(\gamma+2\ln 2-2\sum_{k=1}^{\frac{d}{2}-1}\frac{1}{2k-1}\right)\;,
\end{equation}
when the dimension $d$ is even, while if $d$ is odd, we have
\begin{equation}\label{68a}
\textrm{FP}\,\zeta_{\mathscr{N}}\left(\frac{m}{2}\right)=2\sum_{{\alpha=0 \atop \alpha\neq m-2}}^{d-1}e_{\alpha}\zeta_{H}\left(m-\alpha-1,\frac{d-1}{2}\right)+2e_{m-2}\left(\gamma-\sum_{k=1}^{\frac{d-3}{2}}\frac{1}{2k-1}\right)\;.
\end{equation}
The explicit formulas obtained in (\ref{66a})-(\ref{68a}) are then used in the general results (\ref{45}) and (\ref{60}) in order to obtain the Casimir force
when the piston is a $d$-dimensional sphere.

Before we can continue the analysis of this case, however, we need to pay particular attention to the lowest eigenvalue $\nu=(d-1)/2$ corresponding to the index
$l=0$ in (\ref{61}). In fact, for hybrid boundary conditions of first type the argument of the logarithm in the integrand of (\ref{17b}) behaves, as $\kappa\to 0$, as
\begin{eqnarray}\label{69a}
  I_{\frac{d-1}{2}}(\kappa a)\left[\frac{1-d}{2} K_{\frac{d-1}{2}}(\kappa b)+\kappa bK'_{\frac{d-1}{2}}(\kappa b)\right]-K_{\frac{d-1}{2}}(\kappa a)\left[\frac{1-d}{2} I_{\frac{d-1}{2}}(\kappa b)+\kappa bI'_{\frac{d-1}{2}}(\kappa b)\right]\nonumber\\
  =\left(1-\frac{d}{2}\right)\frac{\Gamma\left(\frac{d-1}{2}\right)}{\Gamma\left(\frac{d+1}{2}\right)}+O(\kappa^{2})\;,
\end{eqnarray}
while for hybrid boundary conditions of second type we have, in region $I$, the following small $\kappa$ expansion
\begin{equation}\label{70}
\left(\beta I_{\frac{d-1}{2}}(a k)+a k I'_{\frac{d-1}{2}}(a k)\right)=\frac{(ak)^{\frac{d+3}{2}}}{2^{\frac{d+1}{2}}\Gamma\left(\frac{d+3}{2}\right)}+O\left(k^{\frac{d+7}{2}}\right)\;.
\end{equation}
Due to the above behavior for $l=0$ as $\kappa\to 0$, the integral representations obtained in (\ref{17b}) and  (\ref{48a}) are not suitable
for the analysis of the lowest eigenvalue $\nu=(d-1)/2$. It is, therefore, necessary to treat the lowest eigenvalue in a different manner
from the higher ones.

Let us start with the case of hybrid boundary conditions of first type. We can write the spectral $\zeta$-function corresponding to the
lowest eigenvalue in region $I$ as follows
\begin{equation}\label{71}
  \zeta_{I}^{\mathcal{H}_{1},\,l=0}(s,a)=\left(\frac{d-1}{2}\right)\frac{\sin(\pi s)}{\pi}\int_{0}^{\infty}d\kappa\,\kappa^{-2s}\frac{\partial}{\partial\kappa}\ln\left[\kappa^{-\frac{d-1}{2}}I_{\frac{d-1}{2}}(\kappa a)\right]\;.
\end{equation}
In order to obtain the analytic continuation at the point $s=-1/2$ we use the asymptotic expansion \cite{gradshtein07}
\begin{equation}\label{72}
  \ln\left[\kappa^{-\frac{d-1}{2}}I_{\frac{d-1}{2}}(\kappa a)\right]=\ln\left[\frac{\kappa^{-\frac{d-1}{2}}\,e^{\kappa a}}{\sqrt{2\pi\kappa a}}\right]+\sum_{n=1}^{\infty}\frac{\mathcal{A}_{n}\left(\frac{d-1}{2}\right)}{(\kappa a)^{n}}\;,
\end{equation}
where the $\mathcal{A}_{n}$ are defined through the relation
\begin{equation}\label{73}
  \sum_{n=1}^{\infty}\frac{\mathcal{A}_{n}\left(\frac{d-1}{2}\right)}{(\kappa a)^{n}}=\ln\left[1+\sum_{j=1}^{\infty}(-1)^{j}\frac{a_{j}\left(\frac{d-1}{2}\right)}{(\kappa a)^{j}}\right]\;,
\end{equation}
 with $a_{0}(z)=1$ and
 \begin{equation}\label{74}
   a_{k}(z)=\frac{1}{n!8^{n}}\prod_{i=1}^{n}\left[4z^2-(2i-1)^2\right]\;.
 \end{equation}
For our purposes of finding the analytic continuation to a neighborhood of $s=-1/2$, it is sufficient to add and subtract
only the first leading term of the expansion (\ref{72}) from the integrand in (\ref{71}). This procedure leads to the result valid for $-1<\Re(s)<1/2$
\begin{eqnarray}\label{75}
  \zeta_{I}^{\mathcal{H}_{1},\,l=0}(s,a)=Z_{I}^{l=0}(s,a)-\left(\frac{d-1}{2}\right)\frac{\sin(\pi s)}{\pi}\left[\frac{d}{2s}-\frac{a}{2s-1}-\frac{d(d-2)}{8a}\frac{1}{2s+1}\right]\;,
\end{eqnarray}
 where we have defined
 \begin{eqnarray}\label{76}
   Z_{I}^{l=0}(s,a)&=&\left(\frac{d-1}{2}\right)\frac{\sin(\pi s)}{\pi}\int_{0}^{\infty}d\kappa\,\kappa^{-2s}\frac{\partial}{\partial\kappa}\Bigg\{\ln\left[\kappa^{-\frac{d-1}{2}}I_{\frac{d-1}{2}}(\kappa a)\right]\nonumber\\
   &-&H(\kappa-1)\Bigg[\ln\left(\frac{\kappa^{-\frac{d}{2}}\,e^{\kappa a}}{\sqrt{2\pi a}}\right)-\frac{1}{\kappa a}\frac{d(d-2)}{8}\Bigg]\Bigg\}\;,
 \end{eqnarray}
 with $H(\kappa-1)$ being the Heaviside step function.

In region $II$ a suitable representation for the spectral $\zeta$-function is
\begin{eqnarray}\label{77}
  \zeta_{II}^{\mathcal{H}_{1},\,l=0}(s,a,b)&=&\left(\frac{d-1}{2}\right)\frac{\sin(\pi s)}{\pi}\int_{0}^{\infty}d\kappa\,\kappa^{-2s}\frac{\partial}{\partial\kappa}
  \ln\Bigg\{K_{\frac{d-1}{2}}(\kappa a)\left[\frac{1-d}{2}I_{\frac{d-1}{2}}(\kappa b)+\kappa b I'_{\frac{d-1}{2}}(\kappa b)\right]\nonumber\\
  &-&I_{\frac{d-1}{2}}(\kappa a)\left[\frac{1-d}{2}K_{\frac{d-1}{2}}(\kappa b)+\kappa b K'_{\frac{d-1}{2}}(\kappa b)\right]\Bigg\}\;.
\end{eqnarray}
To obtain the analytic continuation, an expansion for $\kappa\to\infty$ of the integrand of the above expression is needed. To this end, we exploit the expansion (\ref{72}),
and the following ones
\begin{eqnarray}\label{78}
  I'_{\nu}(z)&\sim& \frac{e^{z}}{\sqrt{2\pi z}}\sum_{n=0}^{\infty}(-1)^{n}\frac{b_{n}(\nu)}{z^{n}}\;,\quad K_{\nu}(z)\sim\sqrt{\frac{\pi}{2z}}e^{-z}\sum_{n=0}^{\infty}(-1)^{n}\frac{a_{n}(\nu)}{z^{n}}\;,\nonumber\\
 K'_{\nu}(z)&\sim&-\sqrt{\frac{\pi}{2z}}e^{-z}\sum_{n=0}^{\infty}(-1)^{n}\frac{b_{n}(\nu)}{z^{n}}\;,
\end{eqnarray}
where the coefficients $b_{n}$ are given by \cite{erdelyi53,gradshtein07}
\begin{equation}\label{79}
b_{0}(x)=1\;,\quad b_{1}(x)=\frac{4x^2+3}{8}\;,\quad b_{n}(x)=\frac{4x^{2}+4n^{2}-1}{n!8^{n}}\prod_{i=2}^{n}\left[4x^2-(2i-3)^2\right]\;.
\end{equation}
By adding and subtracting only the first leading term of the asymptotic expansion of the integrand in (\ref{77}) we obtain the result, valid for $-1<\Re(s)<1/2$,
\begin{eqnarray}\label{80}
  \zeta_{II}^{\mathcal{H}_{1},\,l=0}(s,a,b)&=&Z_{II}^{l=0}(s,a,b)-\left(\frac{d-1}{2}\right)\frac{\sin(\pi s)}{\pi}\Bigg[\frac{a-b}{2s-1}-\frac{1}{2s+1}\left(\frac{3}{8b}+\frac{d-1}{2b}+\frac{(d-1)^{2}}{8b}\right)\nonumber\\
  &+&\frac{1}{2s+1}\frac{d(d-2)}{8a}\Bigg]\;,
\end{eqnarray}
 where we have introduced the function
 \begin{eqnarray}\label{81}
  Z_{II}^{l=0}(s,a,b)&=&\left(\frac{d-1}{2}\right)\frac{\sin(\pi s)}{\pi}\int_{0}^{\infty}d\kappa\,\kappa^{-2s}\frac{\partial}{\partial\kappa}\Bigg\{
  \ln\Bigg[K_{\frac{d-1}{2}}(\kappa a)\left(\frac{1-d}{2}I_{\frac{d-1}{2}}(\kappa b)+\kappa b I'_{\frac{d-1}{2}}(\kappa b)\right)\nonumber\\
  &-&I_{\frac{d-1}{2}}(\kappa a)\left(\frac{1-d}{2}K_{\frac{d-1}{2}}(\kappa b)+\kappa b K'_{\frac{d-1}{2}}(\kappa b)\right)\Bigg]-H(\kappa-1)\Bigg[\ln\left(\sqrt{\frac{b}{a}}\frac{e^{\kappa(b-a)}}{2}\right)\nonumber\\
  &+&\frac{1}{\kappa}\left(\frac{3}{8b}+\frac{d-1}{2b}+\frac{(d-1)^{2}}{8b}-\frac{d(d-2)}{8a}\right)\Bigg]\Bigg\}\;.
 \end{eqnarray}

From the results that we have obtained in region $I$, (\ref{75}), and region $II$, (\ref{81}), we can extract the residue and finite part of the spectral $\zeta$-function
corresponding to the lowest eigenvalue for hybrid boundary conditions of the first kind. More explicitly one has
\begin{equation}\label{82}
  \textrm{FP}\,\zeta_{\mathscr{M}}^{\mathcal{H}_{1},\,l=0}\left(-\frac{1}{2},a,b\right)=Z_{I}^{l=0}\left(-\frac{1}{2},a\right)+Z_{II}^{l=0}\left(-\frac{1}{2},a,b\right)
  -\frac{d-1}{2\pi}\left(\frac{d-b}{2}\right)\;,
\end{equation}
 and
 \begin{equation}\label{83}
   \textrm{Res}\,\zeta_{\mathscr{M}}^{\mathcal{H}_{1},\,l=0}\left(-\frac{1}{2},a,b\right)=-\frac{d-1}{4\pi b}\left(\frac{3}{8}+\frac{d-1}{2}+\frac{(d-1)^{2}}{8}\right)\;.
 \end{equation}
According to (\ref{10b}), and by using the results (\ref{82}) and (\ref{83}), the Casimir force on the piston can be found to have the expression
\begin{equation}\label{84}
  F_{\textrm{Cas}}^{\mathcal{H}_{1},\,l=0}(a)=-\frac{1}{2} Z_{I}^{\prime,\, l=0}\left(-\frac{1}{2},a\right)-\frac{1}{2}Z_{II}^{\prime,\, l=0}\left(-\frac{1}{2},a\right)\;.
\end{equation}

The analysis of the lowest eigenvalue for hybrid boundary conditions of second type follows the same lines that we have described in this section.
Since the general procedure of analytic continuation is, at this point, transparent we present directly the result for the Casimir energy which reads
\begin{equation}\label{85}
  F_{\textrm{Cas}}^{\mathcal{H}_{2},\,l=0}(a)=-\frac{1}{2} W_{I}^{\prime,\, l=0}\left(-\frac{1}{2},a\right)-\frac{1}{2}W_{II}^{\prime,\, l=0}\left(-\frac{1}{2},a\right)\;,
\end{equation}
where the functions $W^{l=0}_{I}$ and $W^{l=0}_{II}$, well defined for $-1<\Re(s)<1/2$, are given by the expressions
\begin{eqnarray}\label{86}
  W_{I}^{l=0}(s,a)&=&\left(\frac{d-1}{2}\right)\frac{\sin(\pi s)}{\pi}\int_{0}^{\infty}d\kappa\,\kappa^{-2s}\frac{\partial}{\partial \kappa}\Bigg\{\ln \left[\kappa^{-\frac{d+3}{2}}\left(\beta I_{\frac{d-1}{2}}( \kappa a)+ \kappa a I'_{\frac{d-1}{2}}( \kappa a)\right)\right]\nonumber\\
&-&H(\kappa-1)\Bigg[\ln\left(\frac{\kappa^{-\frac{d}{2}-1}}{\sqrt{2\pi a}}\,e^{\kappa a}\right)-\frac{1}{\kappa a}\left(\frac{3}{8}+\frac{d-1}{2}+\frac{(d-1)^{2}}{8}\right)\Bigg]\Bigg\}\;,
\end{eqnarray}
and
\begin{eqnarray}\label{87}
  W_{II}^{l=0}(s,a,b)&=&\left(\frac{d-1}{2}\right)\frac{\sin(\pi s)}{\pi}\int_{0}^{\infty}d\kappa\,\kappa^{-2s}\frac{\partial}{\partial\kappa}\Bigg\{
  \ln\Bigg[K_{\frac{d-1}{2}}(\kappa b)\left(\frac{1-d}{2}I_{\frac{d-1}{2}}(\kappa a)+\kappa a I'_{\frac{d-1}{2}}(\kappa a)\right)\nonumber\\
  &-&I_{\frac{d-1}{2}}(\kappa b)\left(\frac{1-d}{2}K_{\frac{d-1}{2}}(\kappa a)+\kappa a K'_{\frac{d-1}{2}}(\kappa a)\right)\Bigg]-H(\kappa-1)\Bigg[\ln\left(\sqrt{\frac{a}{b}}\frac{e^{\kappa(b-a)}}{2}\right)\nonumber\\
  &+&\frac{1}{\kappa}\left(\frac{3}{8a}+\frac{d-1}{2a}+\frac{(d-1)^{2}}{8a}-\frac{d(d-2)}{8b}\right)\Bigg]\Bigg\}\;.
\end{eqnarray}

The total Casimir force on the piston for hybrid boundary conditions of first type is the sum of (\ref{84}) and (\ref{45}) while for
hybrid boundary conditions of second type it is the sum of (\ref{85}) and (\ref{60}) by keeping in mind that
in (\ref{45}) and (\ref{60}) the
lowest angular eigenvalue needs to be omitted. In the next subsections we present explicit results for specific dimensions $d$. We would like to point
out that in the formulas that will follow it is understood that the functions $Z$, $W$ and $\mathscr{F}$ are evaluated for the specific dimension under consideration.
It is important to mention that since the lowest eigenvalue is omitted from the formulas (\ref{45}) and (\ref{60}) the relevant spectral $\zeta$-function becomes now $\bar{\zeta}_{\mathscr{N}}(s)$.
This is related to $\zeta_{\mathscr{N}}(s)$ according to the following relation
\begin{equation}
  \bar{\zeta}_{\mathscr{N}}(s)=\zeta_{\mathscr{N}}(s)-\left(\frac{d-1}{2}\right)^{-2s}\;.
\end{equation}
It is this form of the piston $\zeta$-function that will be used in order to obtain the results of the Casimir force for specific dimensions.

\subsection{Particular Dimensions: Hybrid Boundary Conditions of First Type}

In the following numerical analysis we have set $b=1$.
For $d=2$, thus $D=3$, we have the expression for the force
\begin{eqnarray}
  F_{\textrm{Cas}}^{\mathcal{H}_{1}}(a)&=&-\frac{1}{2}Z'_{I}\left(-\frac{1}{2},a\right)-\frac{1}{2}Z'_{I}\left(-\frac{1}{2},a\right)-\frac{1}{2}\mathscr{F}'_{\mathcal{H}_{1}}\left(-\frac{1}{2},a\right)
  -\frac{1}{2}Z^{\prime,\,l=0}_{I}\left(-\frac{1}{2},a\right)\nonumber\\
  &-&\frac{1}{2}Z^{\prime,\,l=0}_{II}\left(-\frac{1}{2},a\right)+\frac{17}{128a^{2}}\;.
\end{eqnarray}

For $d=3$, namely $D=4$, we have the result
\begin{eqnarray}
  F_{\textrm{Cas}}^{\mathcal{H}_{1}}(a)&=&-\frac{1}{2}Z'_{I}\left(-\frac{1}{2},a\right)-\frac{1}{2}Z'_{I}\left(-\frac{1}{2},a\right)-\frac{1}{2}\mathscr{F}'_{\mathcal{H}_{1}}\left(-\frac{1}{2},a\right)
  -\frac{1}{2}Z^{\prime,\,l=0}_{I}\left(-\frac{1}{2},a\right)\nonumber\\
  &-&\frac{1}{2}Z^{\prime,\,l=0}_{II}\left(-\frac{1}{2},a\right)+\frac{1}{a^{2}}\left(\frac{41297}{163840}+\frac{35}{65536}\gamma+\frac{35}{131072}\ln a^{2}\right)\nonumber\\
  &+&\frac{35}{131072 a^{2}}\left(\frac{1}{\alpha}+\ln\mu^{2}\right)\;.\;\;\;
\end{eqnarray}

For $d=4$, thus $D=5$, we obtain
\begin{eqnarray}
F_{\textrm{Cas}}^{\mathcal{H}_{1}}(a)&=&-\frac{1}{2}Z'_{I}\left(-\frac{1}{2},a\right)-\frac{1}{2}Z'_{I}\left(-\frac{1}{2},a\right)-\frac{1}{2}\mathscr{F}'_{\mathcal{H}_{1}}\left(-\frac{1}{2},a\right)
  -\frac{1}{2}Z^{\prime,\,l=0}_{I}\left(-\frac{1}{2},a\right)\nonumber\\
  &-&\frac{1}{2}Z^{\prime,\,l=0}_{II}\left(-\frac{1}{2},a\right)+\frac{1}{a^{2}}\left(\frac{83485}{221184}-\frac{35}{1572864}\pi^{2}\right)\;.
\end{eqnarray}

And, finally, for $d=5$, thus $D=6$, we get
\begin{eqnarray}
F_{\textrm{Cas}}^{\mathcal{H}_{1}}(a)&=&-\frac{1}{2}Z'_{I}\left(-\frac{1}{2},a\right)-\frac{1}{2}Z'_{I}\left(-\frac{1}{2},a\right)-\frac{1}{2}\mathscr{F}'_{\mathcal{H}_{1}}\left(-\frac{1}{2},a\right)
  -\frac{1}{2}Z^{\prime,\,l=0}_{I}\left(-\frac{1}{2},a\right)\nonumber\\
  &-&\frac{1}{2}Z^{\prime,\,l=0}_{II}\left(-\frac{1}{2},a\right)+\frac{1}{a^{2}}\Bigg(\frac{909728935}{1811939328}-\frac{7285}{25165824}\gamma-\frac{1685}{50331648}\ln a^{2}\nonumber\\
  &+&\frac{565}{25165824}\zeta_{R}(3)\Bigg)-\frac{1685}{50331648 a^{2}}\left(\frac{1}{\alpha}+\ln\mu^{2}\right)\;.
\end{eqnarray}

\begin{figure}[h]
\centering
\mbox{\subfigure[\;$d=2$, and $D=3$]{\includegraphics[width=3in]{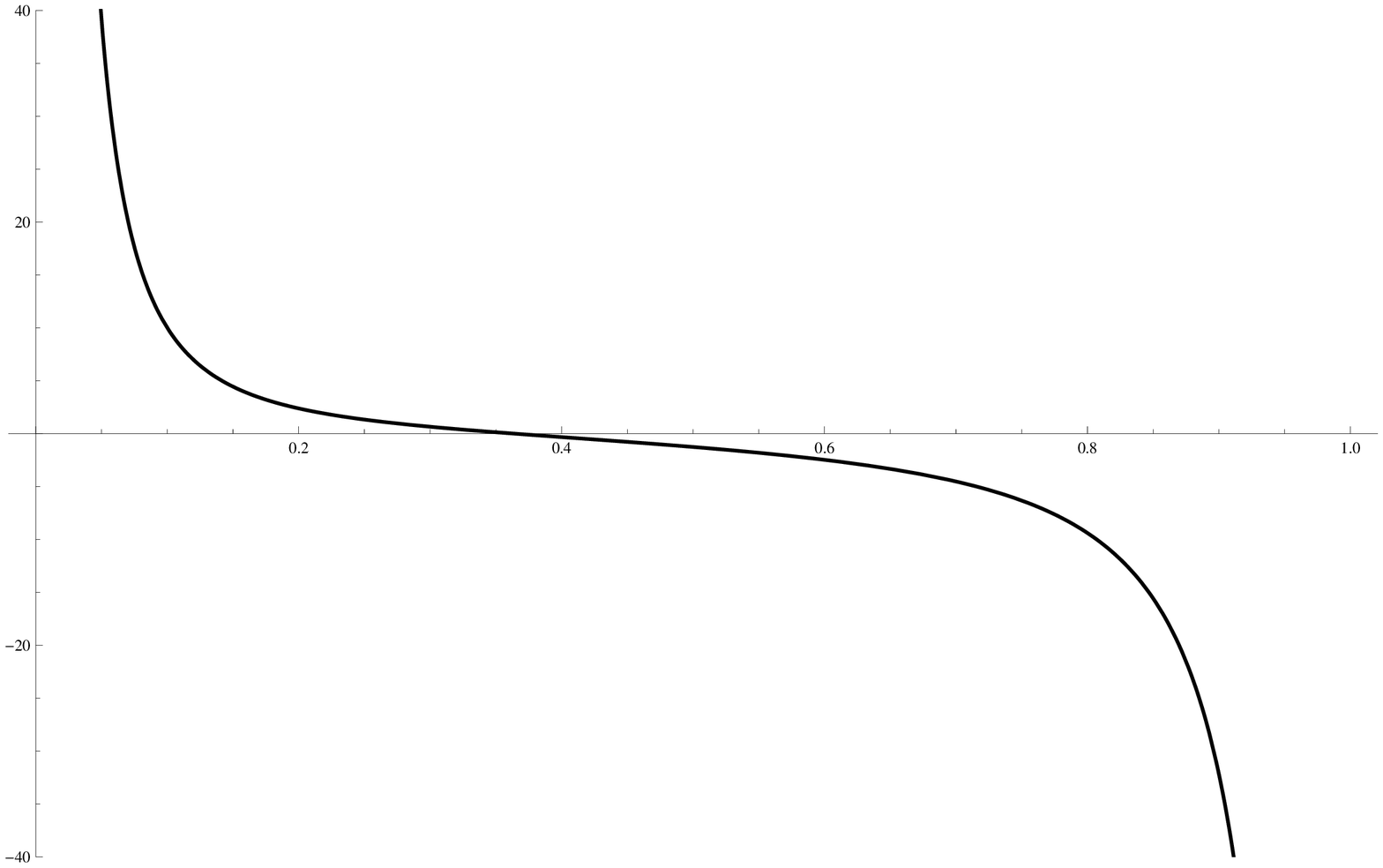}}\quad
\subfigure[\;$d=4$, and $D=5$]{\includegraphics[width=3in]{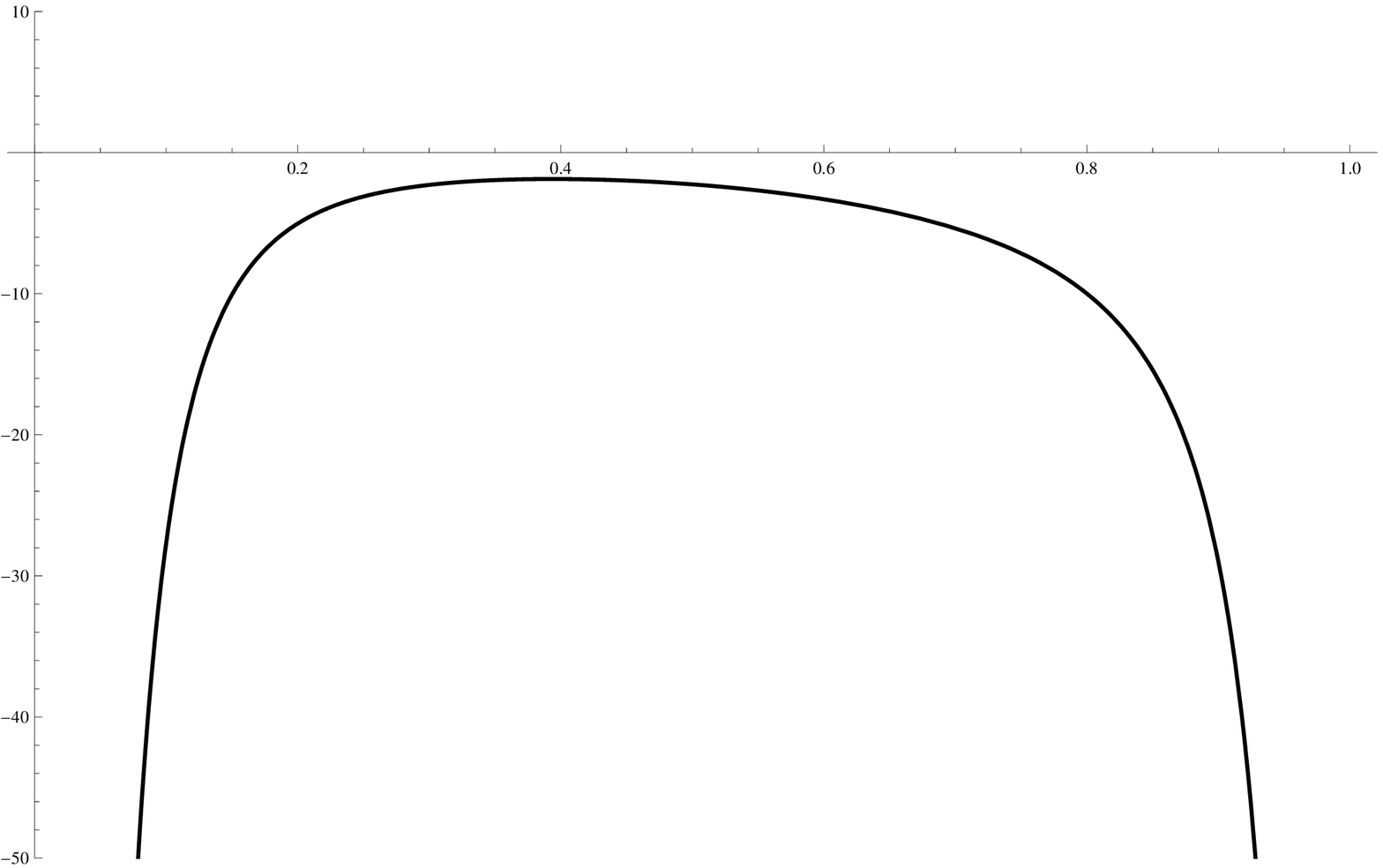} }}
\caption{Plots of the Casimir force, $F^{\mathcal{H}_{1}}_{\textrm{Cas}}(a)$, on the piston $\mathscr{N}$ for hybrid boundary conditions of first type as a function of the position $a$.} \label{fig1}
\end{figure}

It is clear, from the above results, that the Casimir force on the piston $\mathscr{N}$ is not a well defined quantity when $d$ is odd.
The force on the piston for hybrid boundary conditions of first type for $d=2$ and $d=4$ is shown in figure \ref{fig1}.
We can see that for $d=2$ the piston is repelled by both the conical singularity and the base manifold positioned at $r=1$. This means that in this situation, there exists a point of stable
equilibrium. For $d=4$, instead, the piston is attracted by the singularity at $r=0$ and repelled by the base manifold.

\subsection{Particular Dimensions: Hybrid Boundary Conditions of Second Type}

When the piston $\mathscr{N}$ is a sphere of dimension $d=2$, and, therefore, the dimension of $\mathscr{M}$ is $D=3$,
we obtain
\begin{eqnarray}
  F_{\textrm{Cas}}^{\mathcal{H}_{2}}(a)&=&-\frac{1}{2}W'_{I}\left(-\frac{1}{2},a\right)-\frac{1}{2}W'_{I}\left(-\frac{1}{2},a\right)-\frac{1}{2}\mathscr{F}'_{\mathcal{H}_{2}}\left(-\frac{1}{2},a\right)
  -\frac{1}{2}W^{\prime,\,l=0}_{I}\left(-\frac{1}{2},a\right)\nonumber\\
  &-&\frac{1}{2}W^{\prime,\,l=0}_{II}\left(-\frac{1}{2},a\right)+\frac{27}{512a^{2}}\;.
\end{eqnarray}

 \begin{figure}[!]
\centering
\mbox{\subfigure[\;$d=2$, and $D=3$]{\includegraphics[width=3in]{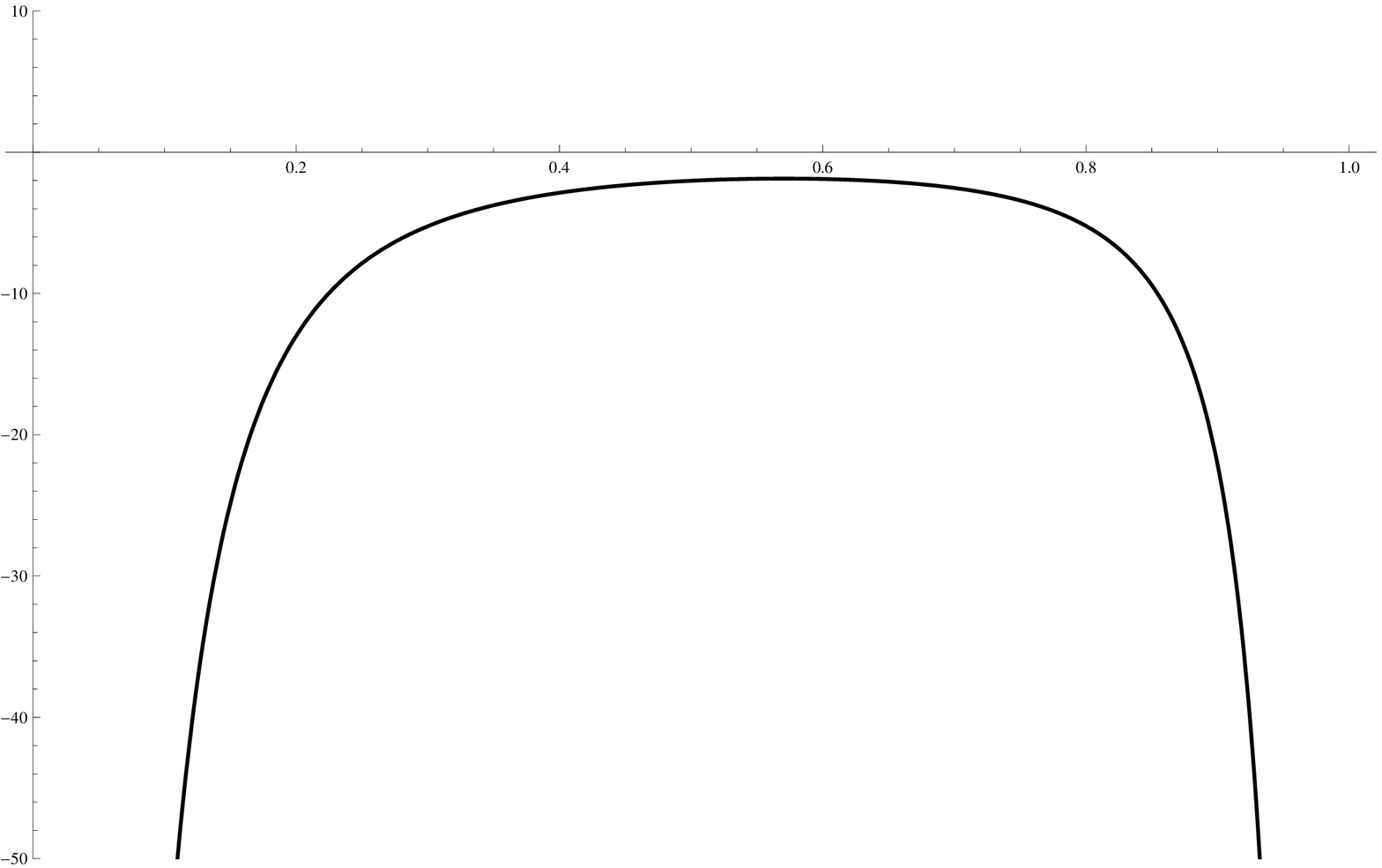}}\quad
\subfigure[\;$d=4$, and $D=5$]{\includegraphics[width=3in]{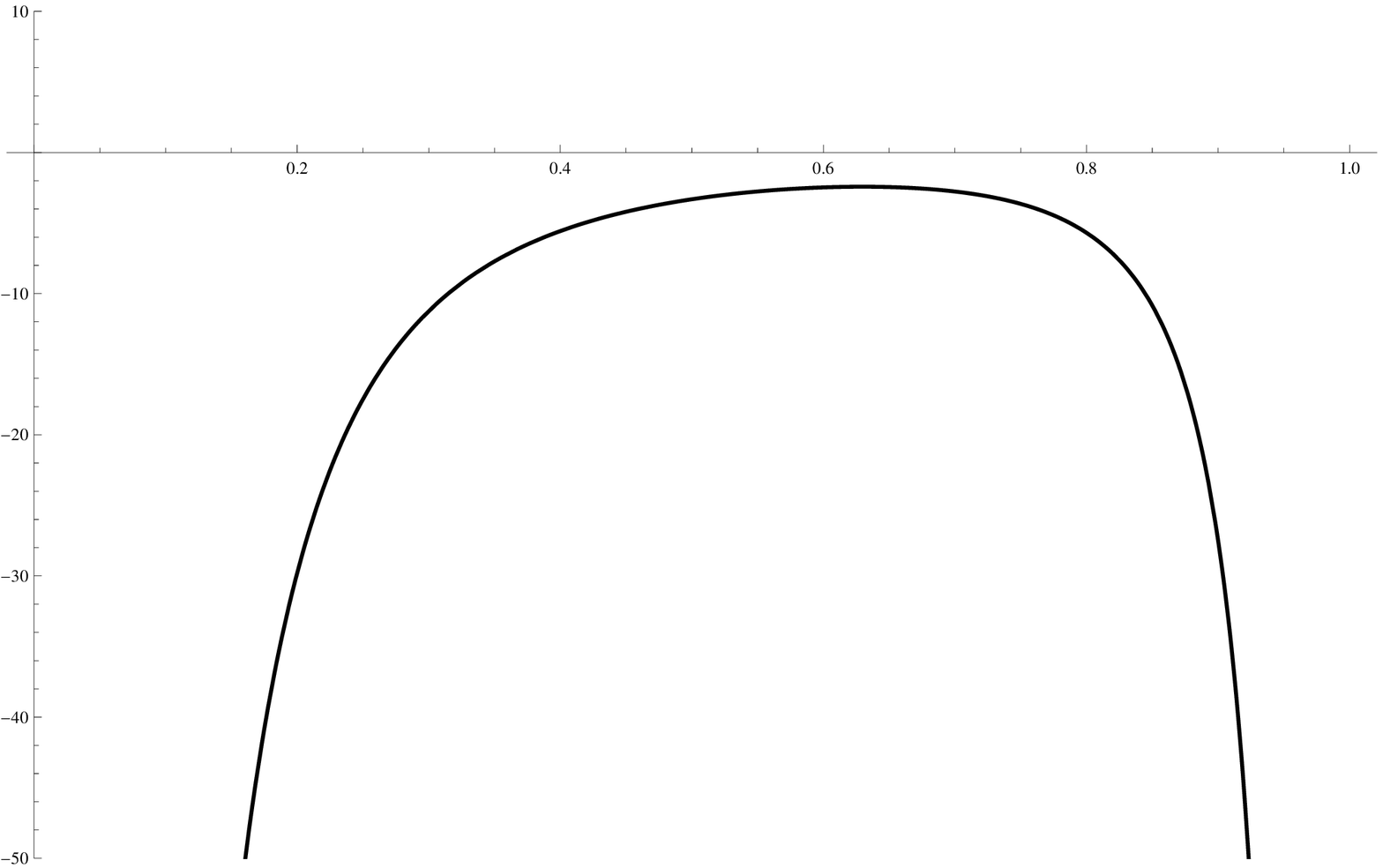} }}
\caption{Plots of the Casimir force, $F^{\mathcal{H}_{2}}_{\textrm{Cas}}(a)$, on the piston $\mathscr{N}$ for hybrid boundary conditions of second type as a function of the position $a$.} \label{fig2}
\end{figure}

For $d=3$, or $D=4$, we have the result
\begin{eqnarray}
  F_{\textrm{Cas}}^{\mathcal{H}_{2}}(a)&=&-\frac{1}{2}W'_{I}\left(-\frac{1}{2},a\right)-\frac{1}{2}W'_{I}\left(-\frac{1}{2},a\right)-\frac{1}{2}\mathscr{F}'_{\mathcal{H}_{2}}\left(-\frac{1}{2},a\right)
  -\frac{1}{2}W^{\prime,\,l=0}_{I}\left(-\frac{1}{2},a\right)\nonumber\\
  &-&\frac{1}{2}W^{\prime,\,l=0}_{II}\left(-\frac{1}{2},a\right)+\frac{1}{a^{2}}\left(\frac{12869}{32768}-\frac{5861}{65536}\gamma-\frac{5861}{131072}\ln a^{2}\right)\nonumber\\
  &-&\frac{5861}{131072 a^{2}}\left(\frac{1}{\alpha}+\ln\mu^{2}\right)\;.\;\;\;\;
\end{eqnarray}

For $d=4$, or $D=5$, we obtain
\begin{eqnarray}
  F_{\textrm{Cas}}^{\mathcal{H}_{2}}(a)&=&-\frac{1}{2}W'_{I}\left(-\frac{1}{2},a\right)-\frac{1}{2}W'_{I}\left(-\frac{1}{2},a\right)-\frac{1}{2}\mathscr{F}'_{\mathcal{H}_{2}}\left(-\frac{1}{2},a\right)
  -\frac{1}{2}W^{\prime,\,l=0}_{I}\left(-\frac{1}{2},a\right)\nonumber\\
  &-&\frac{1}{2}W^{\prime,\,l=0}_{II}\left(-\frac{1}{2},a\right)+\frac{1}{a^{2}}\left(\frac{57781}{221184}+\frac{27253}{1572864}\pi^{2}\right)\;.
\end{eqnarray}

And, finally, for $d=5$, or $D=6$, we get
\begin{eqnarray}
F_{\textrm{Cas}}^{\mathcal{H}_{1}}(a)&=&-\frac{1}{2}Z'_{I}\left(-\frac{1}{2},a\right)-\frac{1}{2}Z'_{I}\left(-\frac{1}{2},a\right)-\frac{1}{2}\mathscr{F}'_{\mathcal{H}_{1}}\left(-\frac{1}{2},a\right)
  -\frac{1}{2}Z^{\prime,\,l=0}_{I}\left(-\frac{1}{2},a\right)\nonumber\\
  &-&\frac{1}{2}Z^{\prime,\,l=0}_{II}\left(-\frac{1}{2},a\right)+\frac{1}{a^{2}}\Bigg(\frac{53466379829}{126835752960}-\frac{1723783}{16777216}\gamma-\frac{1723783}{33554432}\ln a^{2}\nonumber\\
  &+&\frac{10381781}{50331648}\zeta_{R}(3)\Bigg)-\frac{1723783}{33554432 a^{2}}\left(\frac{1}{\alpha}+\ln\mu^{2}\right)\;.
\end{eqnarray}
Once again, the Casimir force retains its ambiguity when the piston $\mathscr{N}$ is odd-dimensional.
The force on the piston for hybrid boundary conditions of second type for $d=2$ and $d=4$ is shown in Figure \ref{fig2}.
We can notice that in both cases, namely $d=2$ and $d=4$, the piston is attracted to the conical singularity at the origin and it is repelled by the
base manifold positioned at $r=1$.

\section{Conclusions}

In this paper we have investigated the Casimir energy and force for massless scalar fields endowed with hybrid boundary conditions in the framework of the conical piston.
This work represents, in particular, a continuation of the studies on conical Casimir pistons initiated in \cite{fucci11}.
By using the methods of $\zeta$-function regularization, we were able to find explicit expressions for the
Casimir energy and force that are valid in any dimension $D$ and for any smooth, compact piston $\mathscr{N}$. The general results are
given in terms of the spectral $\zeta$-function $\zeta_{\mathscr{N}}(s)$ which shows how the Casimir energy and force depend on the
geometry and topology of the piston. In the last section, we have specified our formulas to the case in which
the piston is a $d$-dimensional sphere. In this case $\zeta_{\mathscr{N}}(s)$ can be expressed as a linear combination of Hurwitz $\zeta$-functions, and
we have given explicit numerical results for $d=2$ and $d=4$.

The main interest in considering conical Casimir pistons lies in the fact that the two chambers do not have the same type of geometry.
This feature is particularly relevant in the case of hybrid boundary conditions studied here. For standard Casimir pistons, that are usually considered,
the two chambers possess the same kind of geometry. This means, in particular, that due to the symmetry of the system hybrid boundary conditions
of the type Neumann-Dirichlet and Dirichlet-Neumann yield the same results in the evaluation of the Casimir force.
In the case of conical Casimir pistons this symmetry is not present and, therefore, the two types of boundary conditions lead to essentially
different results for the Casimir energy and for the force acting on the piston. This seems to be a novel result, which has not been observed, due to the reasons explained above,
in standard Casimir pistons with hybrid boundary conditions. These results may also have some relevance in
the framework of field theories with orbifold compactification or in studies of cosmic strings which produce a spacetime that
develops a conical singularity \cite{vilenkin}.

It would be quite interesting to apply the same ideas that led to the construction of the conical piston, to another type of
singular Riemannian manifold termed spherical suspension. Zeta regularization techniques have been exploited in order to
compute the functional determinant for the Laplace operator acting on scalar fields, and the hope is that
a similar investigation might be performed for the evaluation of the Casimir energy for pistons modelled by a spherical suspension \cite{flachi10}.

\begin{acknowledgments}
KK is supported by the National Science Foundation Grant
PHY-0757791.
\end{acknowledgments}

\end{document}